\documentclass[aps,prd,a4paper,nofootinbib,preprintnumbers,superscriptaddress,showpacs,showkeys,twocolumn]{revtex4-2}

\usepackage{natbib}
\usepackage{amsmath}
\usepackage{amssymb}
\usepackage{bm}
\usepackage{graphicx}
\usepackage{subcaption}
\usepackage{color}
\usepackage{slashed}
\usepackage{float}
\usepackage{caption}
\usepackage{graphicx,ifpdf}
\usepackage{upgreek}
\usepackage{floatflt}
\usepackage{hyperref}

\definecolor{darkpastelgreen}{rgb}{0.01, 0.75, 0.24}

\newcommand{\eq}[1]{(\ref{#1})}
\newcommand{\bs}{\boldsymbol}

\def\bbbone{{\mathchoice {\rm 1\mskip-4mu l} {\rm 1\mskip-4mu l} {\rm 1\mskip-4.5mu l} {\rm 1\mskip-5mu l}}}

\usepackage[dvips,textwidth=16.0cm,textheight=24.7cm]{geometry}

\begin{document}
	
    \title{Impact of a hot local CP-odd domain with chirality imbalance on the QCD-biased axionic domain-wall interpretation of NANOGrav 15-year data}

    \author{Zhao Zhang}\email[]{zhaozhang@pku.org.cn}	
	\affiliation{School of Mathematics and Physics, North China Electric Power University, Beijing 102206, China}
    \author{Ruotong Zhao}	
	\affiliation{School of Mathematics and Physics, North China Electric Power University, Beijing 102206, China}

\begin{abstract}
We investigate the influence of the chirality imbalance with local CP-breaking in hot QCD on the generation of a stochastic gravitational wave 
background (SGWB) sourced by the axion-like particle (ALP) domain-wall annihilation, induced by the QCD bias $\Delta V$. Such a bias at finite 
$\theta$ angle, chiral chemical potential $\mu_5$, and temperature $T$ near QCD scale is calculated within a nonlocal Nambu-Jona-Lasinio (NJL) model.
We find that the bias $\Delta V$ decreases as $\theta$ approaches $\pi/2$ from either zero or $\pi$, while increases with rising $\mu_5$. We confirm 
that there exists a $(T,\theta,\mu_5)$ parameter space consistent with the axionic domain-wall interpretation of NANOGrav 15-year data on the nHz 
gravitational waves. We point out that employing the $\theta$-dependent topological susceptibility to quantify the QCD bias at finite $\theta$ 
in the previous study is improper, as it violates the general relation $\Delta V(\theta)=\Delta V(\pi-\theta)$.          

\end{abstract}
	
\date{}
	
\pacs{12.38.Aw,12.38.Mh}

\keywords{Nanohertz gravitational waves, Domain wall, Local CP-breaking, Topological susceptibility, Chiral chemical potential, QCD phase transition}
	
\maketitle
	
\section{Introduction}
	
Gravitational wave (GW) research has opened a new era in the exploration of the mysteries of the Universe. 
In recent years, significant breakthroughs have been achieved in the study of low-frequency (nHz) GW through the pulsar timing arrays (PTA). 
The NANOGrav collaboration, after 15 years of data collection, has reported observations of the SGWB \cite{NANOGrav:2023gor,NANOGrav:2023hvm,NANOGrav:2023hfp}. 
Other recent PTA datasets—including those from the European PTA (EPTA) \cite{EPTA:2023sfo,EPTA:2023akd,EPTA:2023fyk}, the Parkes PTA (PPTA) \cite{Reardon:2023gzh,Reardon:2023zen}, 
and the Chinese PTA (CPTA) \cite{Xu:2023wog}—also provide consistent evidence supporting the existence of nHz stochastic GWs. 

Identifying sources of the SGWB in the nHz band opens up new avenues for revealing the early Universe and exploring the 
new physics beyond the Standard Model (BSM) \cite{NANOGrav:2023hvm,Athron:2023xlk}. Besides the strong candidate of supermassive 
black hole binaries \cite{NANOGrav:2023hfp,Ellis:2023dgf,Shen:2023pan}, the proposed new physics sources include cosmic first-order 
phase transitions \cite{Ashoorioon:2022raz,Tan:2024urn,Han:2023olf,Megias:2023kiy,Fujikura:2023lkn}, 
cosmic strings \cite{Ellis:2020ena,Qiu:2023wbs,Ellis:2023tsl,Wang:2023len,Kitajima:2023vre}, domain walls (DMs)  \cite{Chiang:2020aui,Sakharov:2021dim,King:2023cgv,Kitajima:2023cek}, inflation \cite{Vagnozzi:2023lwo,Borah:2023sbc,Murai:2023gkv,Datta:2023vbs,Chowdhury:2023opo}, scalar-induced GWs \cite{Yi:2023tdk,Wang:2023ost,Liu:2023ymk,Figueroa:2023zhu,Yi:2023mbm}, 
and other astrophysical and cosmological ones \cite{Lambiase:2023pxd,Yang:2023aak,Deng:2023btv,Franciolini:2023wjm,Franciolini:2023pbf}. 
In this work, we focus on the possibility that the nHz range GW is sourced by the DM collapse stemmed from the ALP  
due to the QCD bias. 

Domain walls are the two-dimensional topological defects and a wall serves as a boundary separating regions of space with different degenerate 
vacuum states. In the early Universe, DWs may form due to the spontaneous breaking of the discrete symmetry \cite{Kibble:1976sj}. 
The stable DMs pose a cosmological problem \cite{Zeldovich:1974uw}: their energy density decreases more slowly than that of radiation and 
matter and would eventually overwhelm all other energy components as the Universe expands, leading to a cosmological scenario that is completely 
incompatible with our observed Universe \cite{Barman:2023fad}. 
A mechanism to prevent DMs from dominating the Universe's energy is to introduce a bias potential to break the vacuum degeneracy 
and promote the annihilation of the DM network. 
Since DMs are two-dimensional, their collapse is typically associated with violent and asymmetric spacetime disturbances, leading 
to the generation of GWs.

DWs inherently emerge in models with ALPs, which are generalizations of the Peccei–Quinn axion addressing the strong CP 
problem \cite{Peccei:1977hh,Peccei:1977ur,Wilczek:1977pj} and are widely motivated in the theories BSM \cite{Arvanitaki:2009fg,Holdom:1982ex,Choi:1998ep}.
As light scalars, axion and ALP are also the compelling candidates for dark matter. As a cosmological source of GWs, the 
ALP-DM annihilation induced by a QCD bias has been considered as an attractive scenario \cite{Zhang:2023nrs,Blasi:2023sej,Li:2024psa,Ellis:2023oxs,Lozanov:2023rcd,Gelmini:2023kvo,Geller:2023shn,Bai:2023cqj}, 
which can naturally realize a peak frequency in the nHz band during the QCD phase transition at $T \sim \mathcal{O}(100) \text{MeV}$ \cite{NANOGrav:2023hvm}. 
In this formalism, the QCD-induced bias is provided by the topological susceptibility $\chi_{t}$, a fundamental quantity measuring  
how the QCD vacuum responds to the topological charge fluctuations. The $T$-dependence of $\chi_{t}$ has been investigated 
in lattice QCD simulations \cite{Borsanyi:2016ksw} and the corresponding results have been used in the previous 
calculations \cite{Li:2024psa,Ellis:2023oxs,Lozanov:2023rcd}. Note that the possible CP-odd effect in hot QCD matter 
is ignored in these studies. 

At very high temperature, the sphaleron transition \cite{Manton:1983nd,Klinkhamer:1984di} in the electroweak theory has been 
investigated as a mechanism for baryogenesis \cite{Kuzmin:1985mm,Shaposhnikov:1987tw,Khlebnikov:1988sr,Arnold:1987mh,Arnold:1987zg}. 
Similarly, local CP-odd domains may arise in the hot QCD plasma via strong sphaleron transitions \cite{McLerran:1990de}.
Previous studies suggest that, for high temperature, the gluon configurations with nonzero winding number can be created with 
relatively high probability \cite{McLerran:1990de,Moore:1997im,Moore:1999fs,Moore:2010jd,BarrosoMancha:2022mbj,Bonanno:2023thi}. 
This implies the QCD vacuum characterized by nonzero $\theta$ angle and its fluctuation may become sufficiently sizable and even 
lead to direct experimental signatures \cite{Kharzeev:2007jp,Fukushima:2008xe}. Recently, the impact of a local CP-odd domain in 
hot QCD on the axion DM interpretation of NANOGrav 15-year data (NG15) has been performed  by considering a finite $\theta$ within 
the NJL formalism \cite{Huang:2024nbd}. In this work, the $\theta$-dependent topological susceptibility $\chi_t(\theta)$ is introduced
to evaluate the QCD bias. The main conclusion obtained is that the GW signal strength from the axionic DM collapse due to a QCD 
bias may not be consistent with the NG15 due to the significant suppression of $\chi_{t}(\theta)$ by the $\theta$ parameter 
of order one. In particular, it is found in \cite{Huang:2024nbd} that the magnitude of $\chi_{t}(\theta)$ tends to become significantly 
large at $\theta=\pi$ due to criticality, but the bias leads to too large signal strength to be compatible with the NG15 data.    
    
Note that in the local CP-odd domain of hot QCD, the chirality imbalance emerges naturally due to the axial anomaly. Typically, 
the chiral chemical potential $\mu_5$, which corresponds to the time derivative of $\theta$ 
\footnote{
Note that the spacetime-dependent $\theta(x,t)$ used here is an effective description. Strictly speaking, the fundamental 
QCD $\theta$ is a constant coupling, and the CP-odd domains at finite temperature arise from topological charge fluctuations and 
pseudoscalar excitations (see $\eta^\prime$-like modes or, in beyond standard model contexts, axion/axion-like particle fields).},  
is introduced to quantify the chirality imbalance \cite{Fukushima:2008xe}. It is well known that such an imbalance may lead to 
chiral magnetic effect in heavy-ion collisions \cite{Kharzeev:2007jp,Fukushima:2008xe}. Owing to QCD sphalerons, the chirality 
imbalance may become substantial within the QCD timescale at the early Universe. According to the lattice QCD calculation 
\cite{Astrakhantsev:2019wnp,Braguta:2015owi,Braguta:2015zta}, the chiral condensate (thus also $\chi_{t}$ ) is enhanced significantly 
by a large $\mu_5$. This means the chirality imbalance tends to promote the QCD bias toward the axion DM collapse. Such a catalytic 
effect arising from the chirality imbalance is not taken into account in Ref. \cite{Huang:2024nbd}, where the CP-odd effect is solely 
described by a nonzero $\theta$. 

In this work, we focus on whether the GW signal strengths generated from the axionic DM collapse induced by QCD bias in a local  
CP-odd domain are consistent with the NG15 data by simultaneously considering the nonzero $\mu_5$ and $\theta$. Different from \cite{Huang:2024nbd}, 
a nonlocal two-flavor NJL (NNJL) model with 't Hooft interaction \cite{Hell:2008cc,Pagura:2016pwr,GomezDumm:2005hy,Hell:2009by,Ruggieri:2016ejz,Ruggieri:2020qtq,Blaschke:2007np,Frasca:2011bd,Frasca:2016rsi}
is adopted in our calculation. The main reason for such a choice is that the local NJL used in \cite{Huang:2024nbd} fails to yield 
results qualitatively consistent with the lattice QCD calculation at finite $\mu_5$, whereas the NNJL can do \cite{Ruggieri:2020qtq,Ruggieri:2016ejz}. 
Moreover, the NNJL model better captures non-perturbative features of QCD, such as the nonlocal quark–gluon coupling and confinement-related 
dynamics, which may yield more reasonable results at finite $\theta$, especially at the critical point $\theta=\pi$ for the thermal CP restoration.

As in Refs. \cite{Kitajima:2023vre,Huang:2024nbd}, we assume that the axionic DMs are induced by the spontaneous breaking of $Z_2$ symmetry, where 
a small $Z_2$ breaking potential is a necessary condition for making the DM unstable. We propose that this $Z_2$ breaking potential is induced 
by the coupling between the ALP and gluons \cite{Kitajima:2023vre,Huang:2024nbd} via the axial anomaly during the QCD epoch. Through this approach, 
we can calculate the DM tension and the bias potential; these results allows us to evaluate the signal strengths of generated GWs and compare them 
with the NG15 dataset. In our calculations, the potential bias is obtained via three approaches within the NNJL model: direct evaluation 
of the potential difference, the conventional topological susceptibility, and the $\theta$-dependent topological susceptibility adopted in Ref. \cite{Huang:2024nbd}. 
We argue that, at finite $\theta$, the method used in Ref. \cite{Huang:2024nbd} is improper for calculating the QCD bias.        

The rest of this paper is organized as follows. 
In section \ref{sect:DMQCDbias}, we present the general scenario of the axion DMs and its collapse due to the QCD-induce bias. 
In section \ref{sect:formalizm}, we describe the NNJL model at finite $\theta$ and $\mu_5$ and show the method to calculate the QCD bias and 
the axion domain wall energy density within this formalism. In section \ref{sec:results}, we compare the numerical results with the NANOGrav’s 
15-year observational data and provide relevant discussions. The conclusion and outlook are presented in section \ref{sec:conclusion}.      

\section{Axionic domain walls with QCD bias} \label{sect:DMQCDbias}

\subsection{Axionic domain walls and QCD-bias potential }
    
Following \cite{Kitajima:2023cek,Blasi:2023sej,Li:2024psa,Ellis:2023oxs,Lozanov:2023rcd,Gelmini:2023kvo,Geller:2023shn,Bai:2023cqj}, the ALP $a$ is 
assumed to present prior to the QCD phase transition and a potential  
		\begin{align} 
		V_0(a) = \frac{m_a^2 f_a^2}{n^2} \left(1 - \cos \left(n \frac{a}{f_a} \right) \right) \label{eq:OriginalPt}
	\end{align}
had already been developed by some non-perturbative effects \cite{Huang:2024nbd,Saikawa:2017hiv,Blasi:2023sej,Kitajima:2023cek}. 
Here, $m_a$ and $f_a$ refer to the mass and decay constant of the ALP, respectively, while the integer $n$ denotes the DW number. 
Clearly, the potential \eqref{eq:OriginalPt} satisfies the periodic condition  
    \begin{equation}  
		a/f_a \to a/f_a + 2 \pi 
	\end{equation}
and has degenerate minima at $a/f_a=2\pi k/n$ where $k$ is an integer. This means there are $n$ degenerate distinct vacuum states  
and there exist the DM configurations interpolating between the adjacent vacua $a/f_a=2\pi k/n$ and $2\pi (k+1)/n$. The tension of 
such a DW \cite{Saikawa:2017hiv} is 
\begin{equation} 
\sigma=\frac{8}{n^2}m_af_a^2. \label{eq:tension}
\end{equation}
In the context of the axion models, the potential \eqref{eq:OriginalPt} breaks the $U(1)_{PQ}$ symmetry down to the discrete $Z_n$ 
subgroup, leading to the formation of stable DWs for $n\geq 2$.  
	
As the Universe approaches the QCD scale, an additional potential induced by the $U(1)_A$ anomaly of QCD emerges due to the 
coupling of the field $a$ to gluons through 
\begin{align} 
\mathcal L ~~ \supset ~ &~~ \frac{\alpha_s}{8\pi^2} (n_g \frac{a}{f_a}+\theta) G_{\mu\nu}^a(x) \tilde{G}^{\mu\nu}_{a}(x)\\
&=(n_g \frac{a}{f_a}+\theta) q(x),
\end{align}
where $\alpha_s$ is the strong coupling constant, $G_{\mu\nu}^a(x)$ the gluon field strength tensor, and $\tilde{G}^{\mu\nu,a}(x) = \frac{1}{2} \epsilon_{\mu\nu\rho\sigma} G^{\rho\sigma,a}(x)$ 
its dual. Here, $q(x)$ is the topological charge density, and $n_g$ a factor related to Peccei-Quinn charges of quarks. In this paper, 
the potential $V_{\rm QCD}(a; T, \mu_5,\theta$) can be evaluated within the NNJL formalism. In addition, the cosine 
potential induced by instanton \cite{Gross:1981,GrillidiCortona:2015jxo} with the form  
\begin{align} 
		V_{\rm QCD}(a,\theta) = \chi_{\rm t}\left(1-\cos \left( n_g \frac{a}{f_a} + \theta \right)\right)   
		\,, \label{eq:Biaspt}  
\end{align}
is also adopted and compared with the NNJL potential, where the topological susceptibility $\chi_{\rm t}$ is calculated 
within the NNJL formalism. This type potential is also employed in Ref. \cite{Kitajima:2023cek} to investigate the GWs from the 
QCD-biased domain wall collapse. It is expected that such a simple potential is valid in high temperature where the dilute instanton 
gas is a good approximation. 

The total axion potential then takes the form 
	\begin{align} 
		V_{\rm total}(a) &= V_0(a) + V_{\rm QCD}(a) \nonumber\\ 
    	 &= \frac{m_a^2 f_a^2}{n^2} 
	\left( 1 - \cos\left( n \frac{a}{f_a} \right) \right)+V_{\rm QCD}(a). 
    \end{align}
In this context, $V_{\rm QCD}(a)$ serves as the bias term, which breaks the $Z_n$ symmetry explicitly. For $n \geq 2$, DMs become 
unstable and collapse when the bias is large enough, releasing latent heat into the Universe.
    
Following \cite{Kitajima:2023vre,Huang:2024nbd}, we choose $n=2$ and $n_g=1$. In this case, the vacuum degeneracy lifting 
happens due to the bias potential. Compared to the value at $a/f_a = 0$, the original vacuum at $a/f_a = \pi$ gets the energy shift by 
\begin{align}
  \Delta V &= \Bigm|V_{\text{total}}|_{\frac{a}{f_a}=\pi}-V_{\text{total}}|_{\frac{a}{f_a} = 0}\Bigm| \nonumber\\
  &= \Bigm|V_{\text{QCD}}|_{\frac{a}{f_a}=\pi}-V_{\text{QCD}}|_{\frac{a}{f_a} = 0}\Bigm|. 
  \label{eq:BiasV1}
\end{align}
This non-perturbative bias at finite $T$, $\theta$, and $\mu_5$ near the QCD scale can be calculated using the NNJL model.
Since $V(\theta)$ is an even function of $\theta$ with the period $2\pi$, the QCD potential bias \eqref{eq:BiasV1} satisfies 
the following relation
\begin{eqnarray}
\Delta V (\theta) &=&\Bigm|V_{\text{QCD}}({\pi+\theta})-V_{\text{QCD}}({\theta})\Bigm|\nonumber\\
                  &=&\Bigm|V_{\text{QCD}}({\theta-\pi})-V_{\text{QCD}}({\theta})\Bigm|\nonumber\\
                  &=&\Bigm|V_{\text{QCD}}({\pi-\theta})-V_{\text{QCD}}({\theta})\Bigm|\nonumber\\
                  &=&\Bigm|V_{\text{QCD}}({2\pi-\theta})-V_{\text{QCD}}({\pi-\theta})\Bigm|\nonumber\\
                  &=& \Delta V (\pi-\theta). \label{eq:same2angles}
\end{eqnarray} 
Note that $\theta=\pi/2$ is a special angle because 
\begin{eqnarray}
\Delta V (\pi/2) = 0.
\end{eqnarray}    

If we adopt the cosine potential \eqref{eq:Biaspt}, the bias term \eqref{eq:BiasV1} is simplified as   
\begin{align}
  \Delta V &= |\chi_{\rm t}(T, \mu_5) \bigl(\cos (\pi + \theta) - \cos \theta\bigr)| \nonumber\\
  &=|C(\theta) \chi_{\rm t}(T, \mu_5)|, \label{eq:BiasV2}
\end{align}
where $C(\theta)=2\cos \theta$\footnote{Note that the coefficient $C(\theta)$ is ignored in the corresponding formula of 
the potential bias in Ref. \cite{Huang:2024nbd}.}. 
Note that in Ref. \cite{Huang:2024nbd}, the $\theta$-dependent topological susceptibility $\chi_{\rm t}(T, \theta)$ is introduced and used 
in Eq. \eqref{eq:Biaspt}, which is evaluated in a local NJL model. We stress here that such a replacement is questionable, because the 
topological susceptibility $\chi_{\rm t}$ appearing in the instanton-induced potential \eqref{eq:Biaspt} is defined at $\theta=0$ and therefore 
should be $\theta$-independent. In fact, the general relation \eqref{eq:same2angles}, namely $\Delta V (\theta)=\Delta V (\pi-\theta)$, ceases 
to hold if the $\theta$-independent quantity $\chi_{\rm t}$ in \eqref{eq:BiasV2} is substituted by $\chi_{\rm t}(\theta)$. In our numerical 
calculations, we point out that this substitution can yield misleading conclusions, as demonstrated by the spurious sharp enhancement 
of the QCD-induced potential bias at $\theta=\pi$.

The DMs begin to collapse when the potential bias $\Delta V$ is comparable to the DM energy density $\rho_{\text{DW}}(T, \mu_5,\theta)$. 
This means that the maximal amount of the released latent heat corresponds to 
	\begin{align} 
		\rho_{\rm DW} \sim \Delta V(T,\mu_5, \theta)\,.  
	\end{align}
In the following, we will adopt the numerical analysis result
\begin{align} 
\rho_{\rm DW} \simeq 0.5 \Delta V \label{eq:rouDW}
\end{align} 
obtained in Refs. \cite{Saikawa:2017hiv,Hiramatsu:2010yz}. 

\subsection{GWs from DW network collapse}
	
We assume that the axionic DM annihilation due to the QCD bias happens at $T=T_*$  and is the only source of the GW. The 
produced GW power spectrum at the peak frequency is then assessed 
via the following signal strength \cite{Blasi:2023sej}
	\begin{align} 
	\alpha_*(T_*,\mu_5, \theta) 
	&= 
	\frac{\rho_{\rm DW}(T_*,\mu_5, \theta)}{\rho_{\rm rad}(T_*)}\simeq  
	\frac{0.5 \Delta V(T_*, \mu_5,\theta)}{\rho_{\rm rad}(T_*)} 
	\,,  \label{alpha-star1}
	\end{align}
where $g_*(T_*)$ is the effective number of relativistic degrees of freedom at $T=T_*$ and 
\begin{equation}
\rho_{\rm rad}(T) = \frac{\pi^2}{30}g_*(T) T^4. 
\end{equation}
As in Ref. \cite{Huang:2024nbd}, the effective degrees of freedom $g_*$(T) near the QCD phase transition temperature can be taken 
from Ref. \cite{ParticleDataGroup:2024cfk}. If we use the simple potential bias \eqref{eq:BiasV2}, the signal strength reads 
\begin{align} 
	\alpha_*(T_*,\mu_5, \theta) 
	&\simeq 0.15 
	\times 
	\left( \frac{|C(\theta)\chi_{\rm t}(T_*,\mu_5)|^{1/4}}{100\,{\rm MeV}} \right)^4\notag\\
	&\times\left( \frac{T_*}{ 100\,{\rm MeV}} \right)^{-4}
	\left( \frac{g_*(T_*)}{10} \right)^{-1}
	\,.  \label{alpha-star2}
	\end{align}

We compute the signal strength $\alpha_*$ using three methods within the NNJL model:
(1) direct calculation of the potential bias at the mean-field level; (2) adopting Eq. \eqref{alpha-star2} and evaluating the topological 
susceptibility $\chi_t(T,\mu_5)$; (3) following Ref. \cite{Huang:2024nbd} and substituting $\chi_t(T,\mu_5)$ in Eq. \eqref{alpha-star2} 
with the generalized $\theta$-dependent topological susceptibility $\chi_t(T,\mu_5,\theta)$. We then compare these results with the NG15 
dataset. For convenience, these three approaches are hereafter referred to as Method I, Method II, and Method III, respectively.   

\section{NNJL model analysis of the potential bias and axionic DW energy density at nonzero $\theta$ and $\mu_5$} \label{sect:formalizm}	

\subsection{The formalism}

The non-perturbative formalism we used to calculate the QCD bias is the type of two flavor NNJL model, which has been extensively 
used to investigate the QCD phase transition at finite temperature and baryon density \cite{Hell:2008cc,Pagura:2016pwr,GomezDumm:2005hy,Ruggieri:2020qtq,Ruggieri:2016ejz,Hell:2009by,Blaschke:2007np,Frasca:2011bd,Frasca:2016rsi}.   	
In the QCD Lagrangian, the topological angle $\theta$ appears as a CP-odd term  
	\begin{equation}
		\delta {\mathcal L}_\theta = \theta q(x)=\theta\frac{g^2}{32\pi^2} G_{\mu\nu}^a(x) \tilde{G}^{\mu\nu,a}(x),  
		\label{L:theta}
	\end{equation}
where $G_{\mu\nu}^a(x)$ is the gluon field strength tensor and $\tilde{G}^{\mu\nu,a}(x) = \frac{1}{2} \epsilon_{\mu\nu\rho\sigma} G^{\rho\sigma,a}(x)$ 
is its dual, and $q(x)$ is the topological charge density. In the NJL formalism, the angle $\theta$ can be introduced via the instanton induced quark 
interactions. 

We adopt the two flavor NNJL model used in Ref. \cite{Ruggieri:2020qtq},  which Lagrangian density includes three terms  
	\begin{equation}
		{\cal L} = {\cal L}_q + {\cal L}_m +{\cal L}_4. 
		\label{eq:L1}
	\end{equation}
The first is  
	\begin{equation}
		{\cal L}_q =\bar\psi(i\gamma^\mu\partial_\mu   +\mu_5 \gamma^0\gamma^5)\psi, 
		\label{Lq}
	\end{equation}
which denotes the free quark contribution with the chiral chemical potential $\mu_5$. The second is the mass term  
	\begin{equation}
		{\cal L}_m = -m_0\bar\Psi\Psi,
		\label{L:m}
	\end{equation}
where $m_0$ is the current quark mass. Note that the dressed quark filed  
	\begin{equation}
		\Psi(x) = \int~d^4y~G(x-y)\psi(y)
		\label{Psi}
	\end{equation}
is introduced in \eqref{L:m}, where $G(x-y)$ denotes an formfactor. Using this special form of \eqref{L:m} rather than the 
standard one $\mathcal{L}_m^{(0)}= -m_0 \bar\psi \psi$ aims to mimic the perturbative tail of the current quark mass at large 
Euclidean momentum \cite{Ruggieri:2020qtq}.
The last is the term of nonlocal four-quark interactions 
	\begin{align}
		{\cal L}_4 &= G_1\sum_{\ell=0}^3\left[(\bar Q \tau_\ell Q)^2 + (\bar Q i\gamma^5 \tau_\ell Q)^2\right] \notag \\
		&\quad + 8G_2 \left[e^{i\theta}\det(\bar Q_R Q_L) + e^{-i\theta}\det(\bar Q_L Q_R)\right],
		\label{L4}
	\end{align}
where $\tau_l = (\bbbone, i{\bs \tau})$ is a quaternion in flavor space and the spinor $Q$ denotes another dressed quark field
	\begin{equation}
		Q(x) = \int~d^4y~F(x-y)\psi(y),
		\label{form_factor}
	\end{equation}
defined via the form factor $F(x-y)$. The first part of the interactions \eqref{L4} represents the nonlocal coupling between 
scalar (pseudoscalar) bilinears of the quark fields, which respects chiral symmetry; The second is the nonlocal 't Hooft
interaction induced by QCD instantons, which violates the $U(1)_A$ symmetry. As mentioned, the nonzero angle $\theta$ appears 
in this term,  indicating the breaking of strong CP symmetry.         

The form factors $G(x-y)$  and  $F(x-y)$ in \eqref{Psi} and \eqref{form_factor} will be specified later. Clearly, the local NJL
corresponds to the case with $F(x-y)=\delta(x-y)$, which has been adopted to study properties of QCD axion in Ref. \cite{Lu:2018ukl}.
In all forms, we have $G_1 = (1 - c)G$, $G_2 = cG$ and $c = 0.2$ \cite{Lu:2018ukl}.

The explicit $\theta$ dependence in \eqref{L4} can be removed by performing the chiral rotation
	\begin{equation}
		\psi_R \rightarrow e^{-i\theta/4}\psi_R\equiv \psi'_R,~\psi_L \rightarrow e^{i\theta/4}\psi_L \equiv \psi'_L,
	\end{equation}
and the Lagrangian (\ref{L4}) turns into 
	\begin{eqnarray}
		{\cal L'}_4 &=&G_1\sum_{\ell=0}^3\left[(\bar Q'  \tau_\ell Q')^2 +
		(\bar Q'  i\gamma^5 \tau_\ell Q')^2\right]\nonumber\\
		&&+8G_2 \left[ \mathrm{det}(\bar Q'_R  Q'_L)
		+ \mathrm{det} (\bar Q'_L  Q'_R)\right]\label{eq:L42}
	\end{eqnarray}
with 
\begin{equation}
		Q'(x) = \int~d^4y~F(x-y)\psi'(y).
		\label{form_factor2}
	\end{equation}
Consequently, the $\theta$ parameter appears in the quadratic part of the Lagrangian (\ref{eq:L1}) and the mass term becomes
	\begin{eqnarray}
		{\cal L'}_m = -\bar\Psi'(  m_{0+} +i m_{0-}\gamma^5  )\Psi',
		\label{eq:Lq2}
	\end{eqnarray}
where 
	\begin{eqnarray}
       \Psi'(x) = \int~d^4y~G(x-y)\psi'(y),~~~~~~~~\\
		m_{0+} =  m_0 \cos(\theta/2), ~~ m_{0-} =  m_0 \sin(\theta/2).
	\end{eqnarray}
	
Following Ref. \cite{Ruggieri:2020qtq}, we also introduce the collective fields in the primed bilinear forms   
	\begin{eqnarray}
		\sigma' &=& G_+\bar Q' Q',\\
		\eta' &=& G_-\bar Q'i\gamma^5 Q',
		\label{eq:eta}
	\end{eqnarray}
where $G_\pm = G_1 \pm G_2$. The corresponding original bilinear fields can be obtained via the orthogonal rotation
\begin{eqnarray}
\begin{pmatrix} \bar Q Q \\\bar Qi\gamma^5 Q \end{pmatrix}=\begin{pmatrix} \cos\frac{\theta}{2} & \sin\frac{\theta}{2} \\ -\sin\frac{\theta}{2} & \cos\frac{\theta}{2} \end{pmatrix} \begin{pmatrix} \bar Q' Q' \\\bar Q'i\gamma^5 Q' \end{pmatrix}.		
\end{eqnarray}
We use $\sigma'_0$ ($\eta'_0$) to denote the vacuum condensate of the field $\sigma' (\eta')$.
The mean field thermodynamic potential at the one-loop approximation for finite $T$, $\mu_5$, and $\theta$ takes the form
    \begin{equation}
    	\begin{split}
    		\Omega_\text{M} &= \frac{{\sigma'}_0^2}{G_+} + \frac{{\eta'}_0^2}{G_-} \label{eq:omega_1} \\
    		&\quad - 2 N_c T \sum_n \int \frac{d^3p}{(2\pi)^3} \log \left[ \beta^4 \left(\omega_n^2 + E_+^2\right) \right. \\
    		&\quad \left. \times \left(\omega_n^2 + E_-^2\right) \right], 
    	\end{split} 
    \end{equation}
where $\beta=1/T$ and $\omega_n=(2n+1)\pi T$ are the Matsubara frequencies for quark fields. The energy branches $E_{\pm}$ in Eq.~\eq{eq:omega_1} 
are defined as 
	\begin{equation}
		E^2_\pm(p)  = (|\bm p|\pm\mu_5)^2 + {\cal M}^2(p) + {\cal N}^2(p),
		\label{eq:E2:pm}
	\end{equation}
where $p \equiv |\boldsymbol{p}|$ and the two momentum dependent masses read
	\begin{eqnarray}
		{\cal M}(p) &=& m_{0+}{\cal R}(p)  -2{\cal C}(p)\sigma'_0,\label{eq:mms:M}\\
		{\cal N}(p) &=& m_{0-}{\cal R}(p)  -2{\cal C}(p)\eta'_0.\label{eq:mms:N}
	\end{eqnarray}
The functions ${\cal C}(p)$ and ${\cal R}(p)$ are defined as
\begin{eqnarray}
		{\cal C}(p)\equiv F^2(p),\label{eq:C}\\
		{\cal R}(p)\equiv G^2(p),\label{eq:R}
\end{eqnarray}
where F(p) (G(p)) is the Fourier transform of the form factor $F(x-y)(G(x-y))$. 

The condensates $\sigma'_0$ and $\eta'_0$ are determined by the stationary condition
\begin{eqnarray}
		\frac{\partial {\Omega_\text{M}}}{\partial {\sigma'_0}}=\frac{\partial {\Omega_\text{M}}}{\partial {\eta'_0}}=0. \label{eq:GapEqs}
\end{eqnarray}
The related original two condensates can be obtained by the transition
\begin{eqnarray}
\begin{pmatrix} \langle\bar Q Q\rangle \\ \langle\bar Qi\gamma^5 Q\rangle \end{pmatrix}=\begin{pmatrix} \cos\frac{\theta}{2} & \sin\frac{\theta}{2} \\ -\sin\frac{\theta}{2} & \cos\frac{\theta}{2} \end{pmatrix} \begin{pmatrix} \sigma'_0/G_+ \\\eta'_0/G_- \end{pmatrix}.		
\end{eqnarray}

\vspace{5pt}
\begin{table*}[ht]
		\centering
		\caption{Model parameters}
		\begin{tabular}{lccccc}
			\hline
			  ~~ & $m_0$ [MeV]&~~~~ $\Lambda$ [MeV]  & $G$ & $\langle \bar{\psi} \psi \rangle$ [$(\mathrm{MeV})^3$] &  ~~~~ $\Lambda_{\text{UV}}$ [MeV]\\
			\hline
			   & 5    & 550  & $2.6/\Lambda^2$     & $(-250)^3$  &  $3\Lambda$\\
			\hline
		\end{tabular}
		\label{tab:parameters}
\end{table*}
\vspace{5pt}

\subsection{The model parameters}

In this paper, we utilize the analytical forms of ${\cal R}(p)$ and ${\cal C}(p)$ used in the previous studies \cite{Ruggieri:2020qtq,Ruggieri:2016ejz}, which reads  
	
\begin{eqnarray}
		{\cal C}(p_E) &=& \theta(\Lambda^2 - p_E^2) \nonumber\\
		&+& \theta(p_E^2 -\Lambda^2) \frac{\Lambda^2}{p_E^2}
		\frac{\left(\log \Lambda^2/\Lambda^2_\mathrm{QCD}\right)^\gamma}
		{\left(\log p_E^2/\Lambda^2_\mathrm{QCD} \right)^{\gamma}}\nonumber\\
		&&\label{eq:CPE}
	\end{eqnarray}
and
	\begin{eqnarray}
		{\cal R}(p_E) &=& \theta(\Lambda^2 - p_E^2) \nonumber\\
		&+& \theta(p_E^2 -\Lambda^2)
		\frac{\left(\log \Lambda^2/\Lambda^2_\mathrm{QCD}\right)^{d_m}}
		{\left(\log p_E^2/\Lambda^2_\mathrm{QCD} \right)^{d_m}}.\nonumber\\
		&&\label{eq:RPE}
	\end{eqnarray}
Here $p_E$ refers to the Euclidean 4-momentum and $\gamma=1-d_m$. The parameter $d_m = 12/29$ is the anomalous dimension of the current quark mass for a two-flavor QCD.
	 
The effective momentum cutoff parameter $\Lambda$ in ${\cal C}(p)$ is fixed by reproducing the vacuum quark condensate 
and pion decay constant in combination with the coupling $G$ and current quark mass $m_0$. The corresponding parameters $m_0$, $G$, and $\Lambda$ are listed in Table 
\ref{tab:parameters} and are adopted from Ref. \cite{Ruggieri:2020qtq}.

\begin{figure*}[!htbp]
	\centering
	\begin{subfigure}{0.48\textwidth}
     \includegraphics[width=\linewidth]{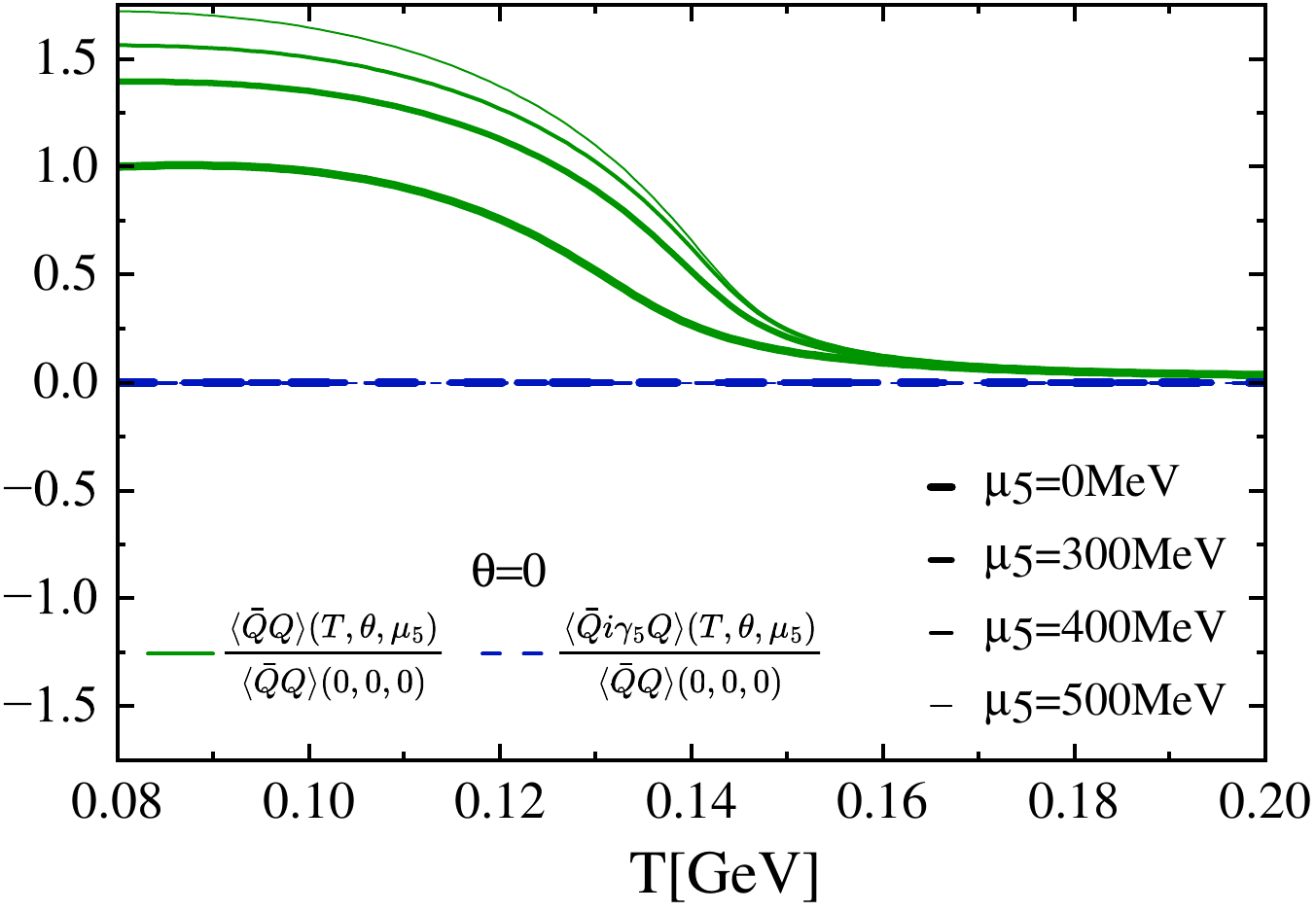}
     \subcaption{}
     \label{fg:fg1-1}
     \end{subfigure}
  \hfill
     \begin{subfigure}{0.48\textwidth}
     \includegraphics[width=\linewidth]{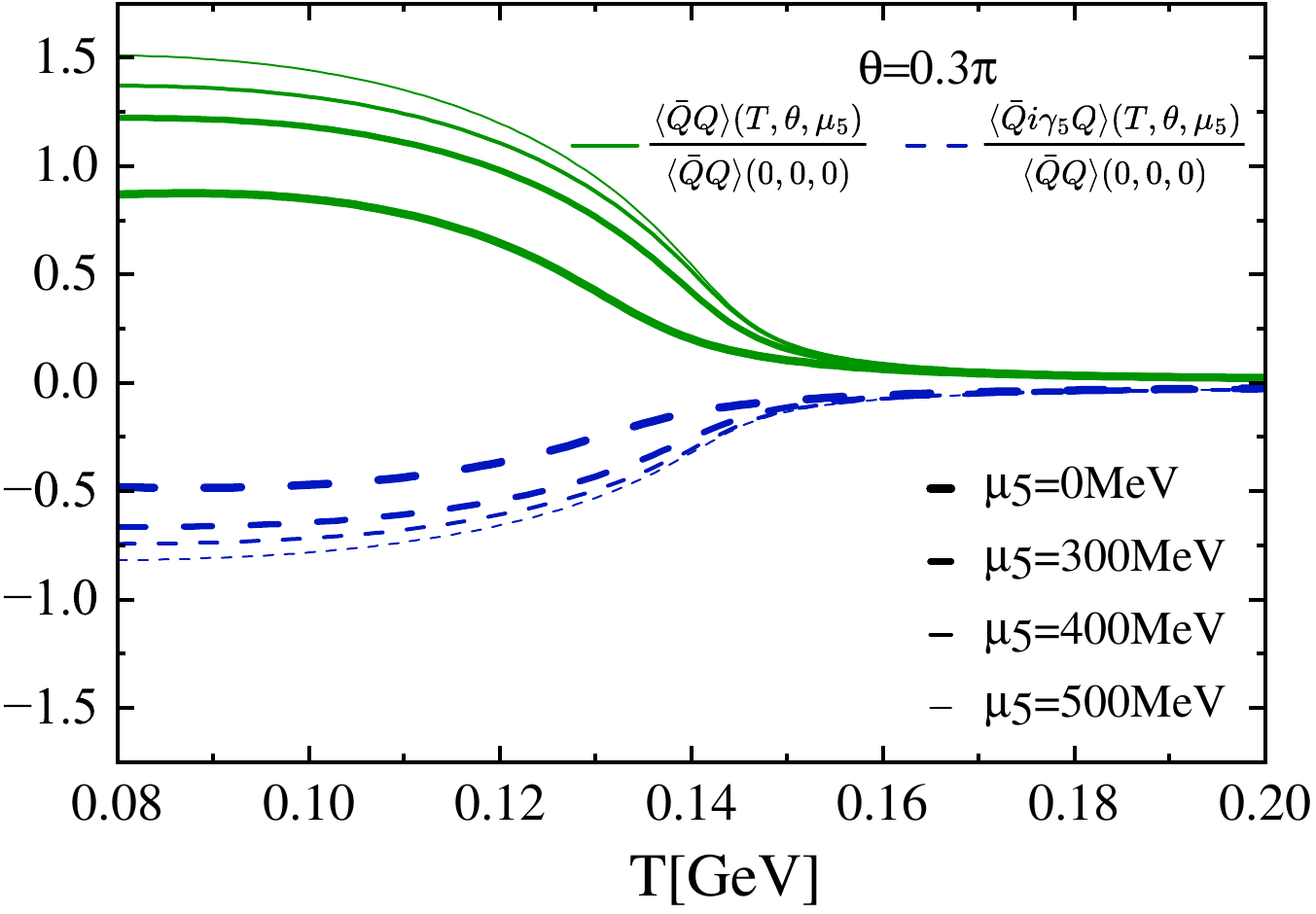}
     \subcaption{}
     \label{fg:fg1-2}
    \end{subfigure}
  \hfill  
    \begin{subfigure}{0.48\textwidth}
     \includegraphics[width=\linewidth]{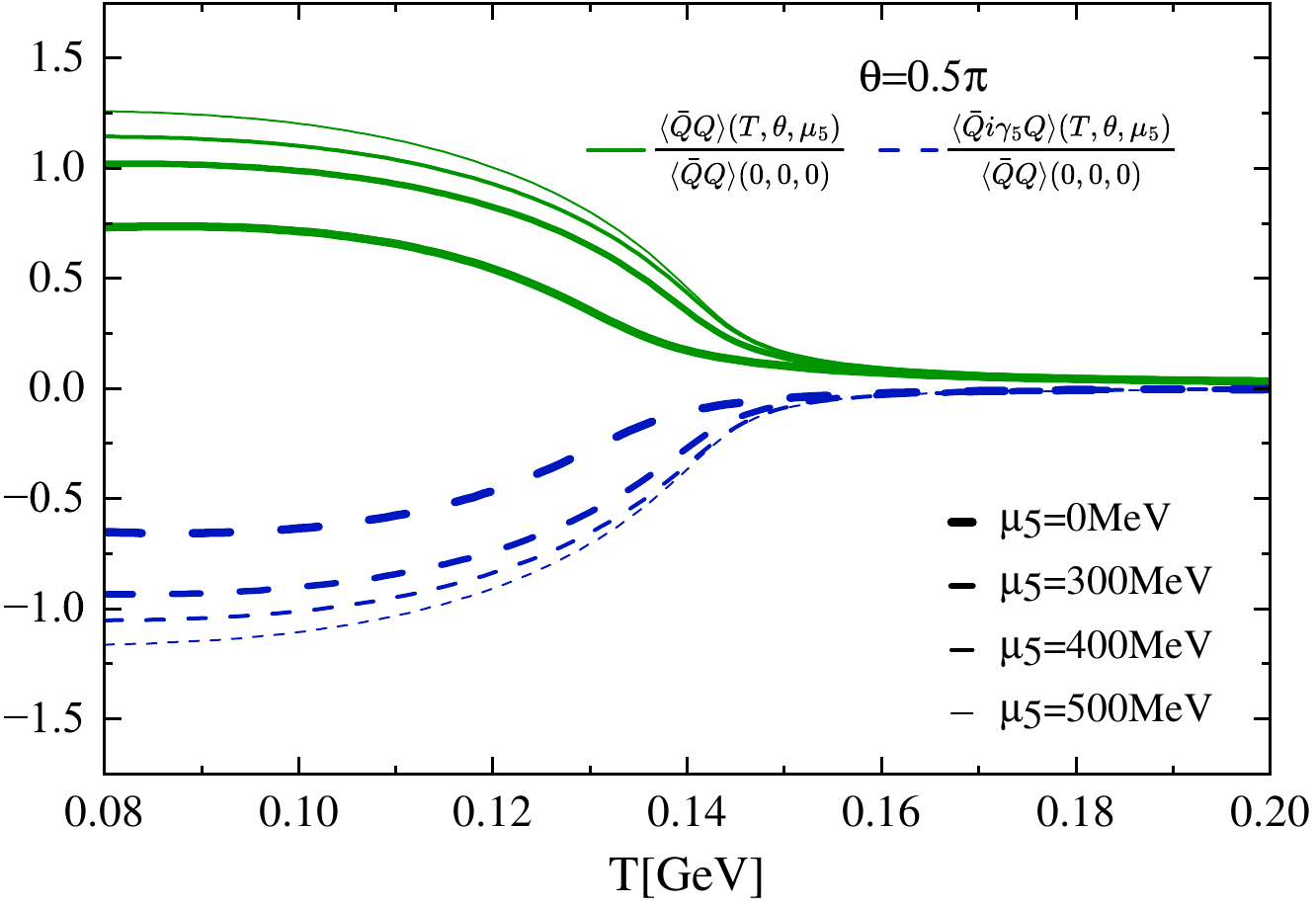}
     \subcaption{}
     \label{fg:fg1-3}
    \end{subfigure}
  \hfill
   \begin{subfigure}{0.48\textwidth}
     \includegraphics[width=\linewidth]{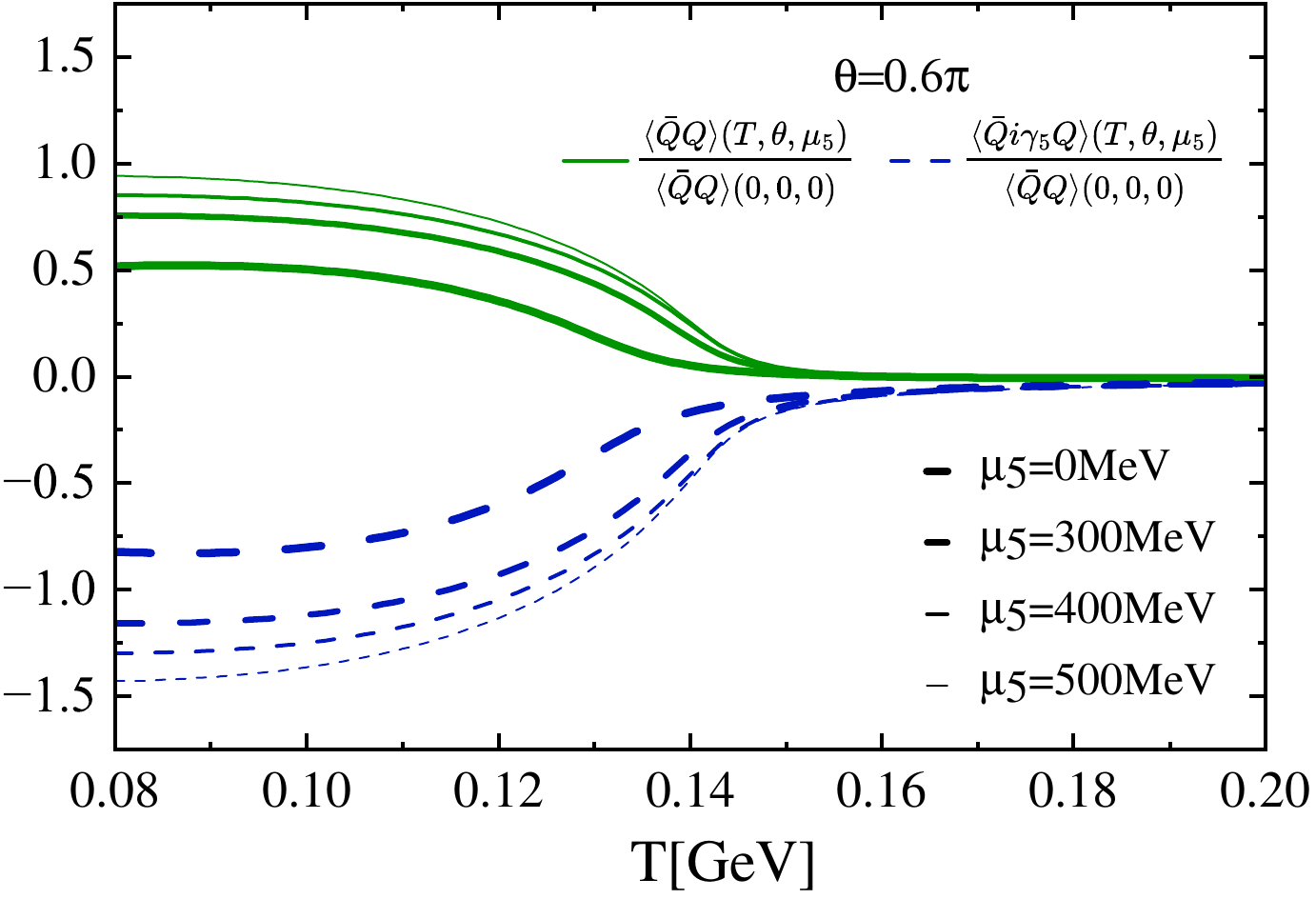}
     \subcaption{}
     \label{fg:fg1-4}
    \end{subfigure}
  \hfill  
    \begin{subfigure}{0.48\textwidth}
     \includegraphics[width=\linewidth]{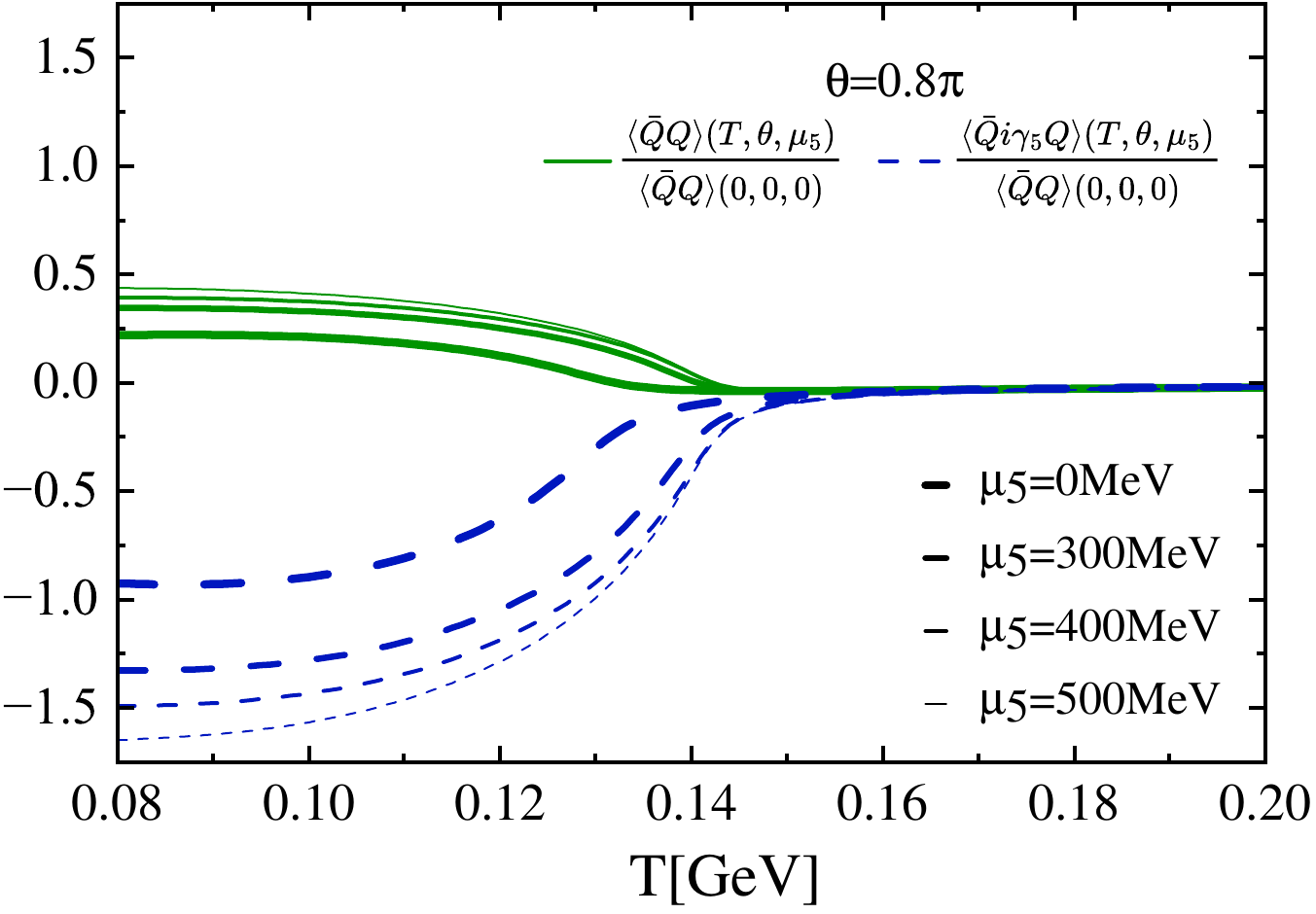}
    \subcaption{}
    \label{fg:fg1-5}
    \end{subfigure} 
  \hfill  
    \begin{subfigure}{0.48\textwidth}
     \includegraphics[width=\linewidth]{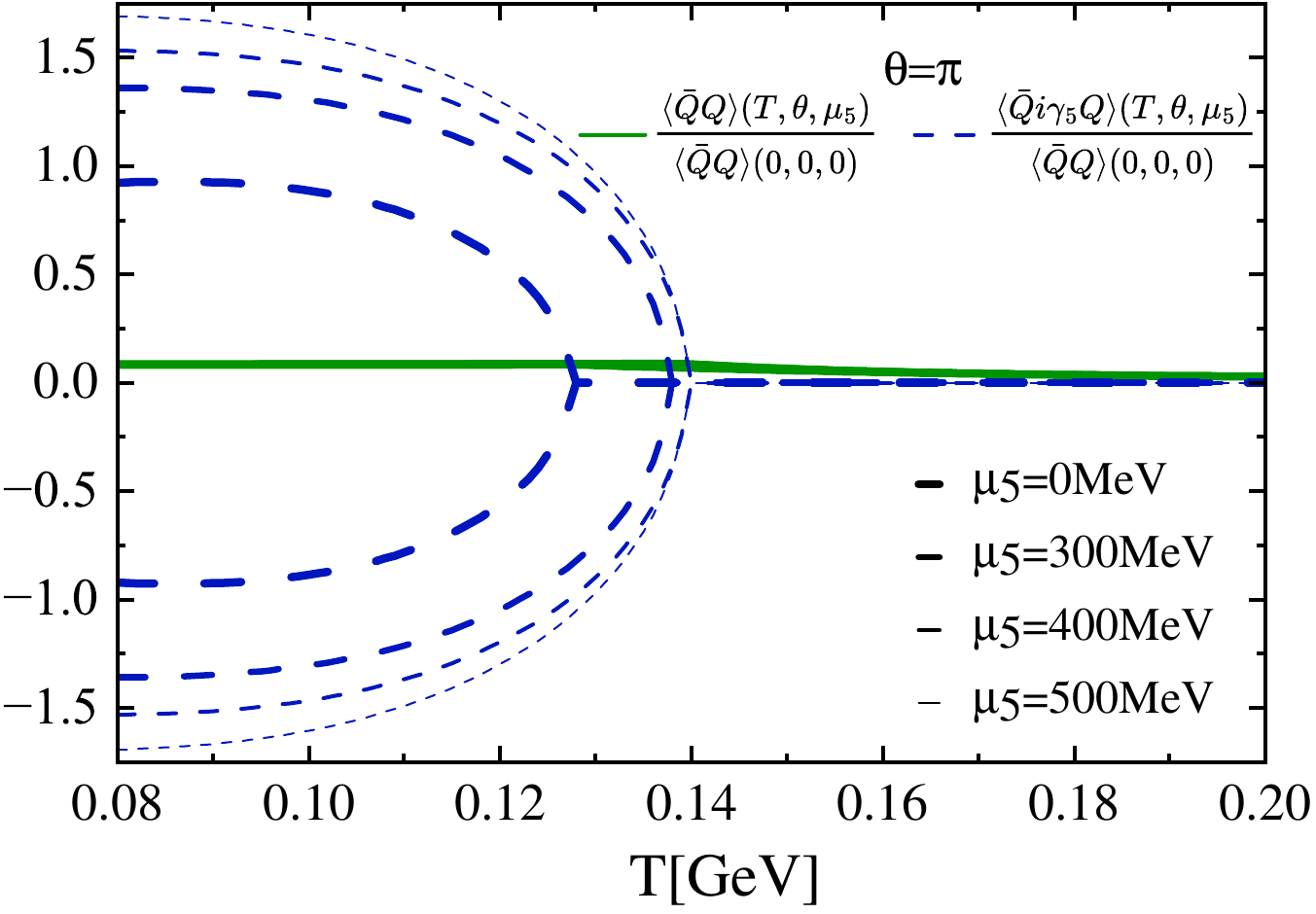}
     \subcaption{}
     \label{fg:fg1-6}
    \end{subfigure}
	\caption{ Normalized scalar and pseudoscalar condensates versus temperature for different $\mu_5$ and $\theta$.}
	\label{fg:condensates1}
\end{figure*}

\begin{figure*}[!htbp]
	\centering
	\begin{subfigure}{0.48\textwidth}
     \includegraphics[width=\linewidth]{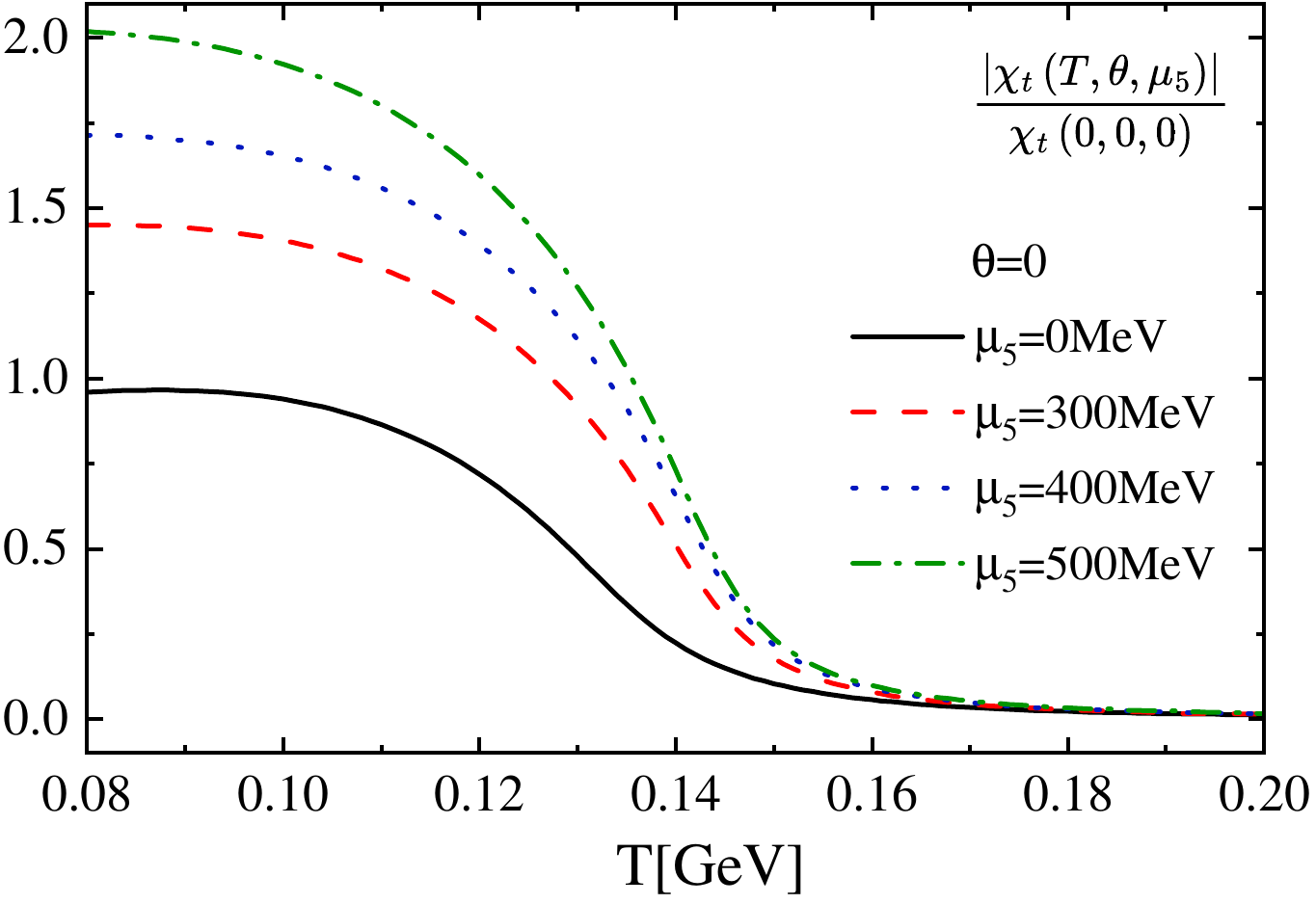}
     \subcaption{}
      \label{fg:fg2-1}
    \end{subfigure}
  \hfill
     \begin{subfigure}{0.48\textwidth}
     \includegraphics[width=\linewidth]{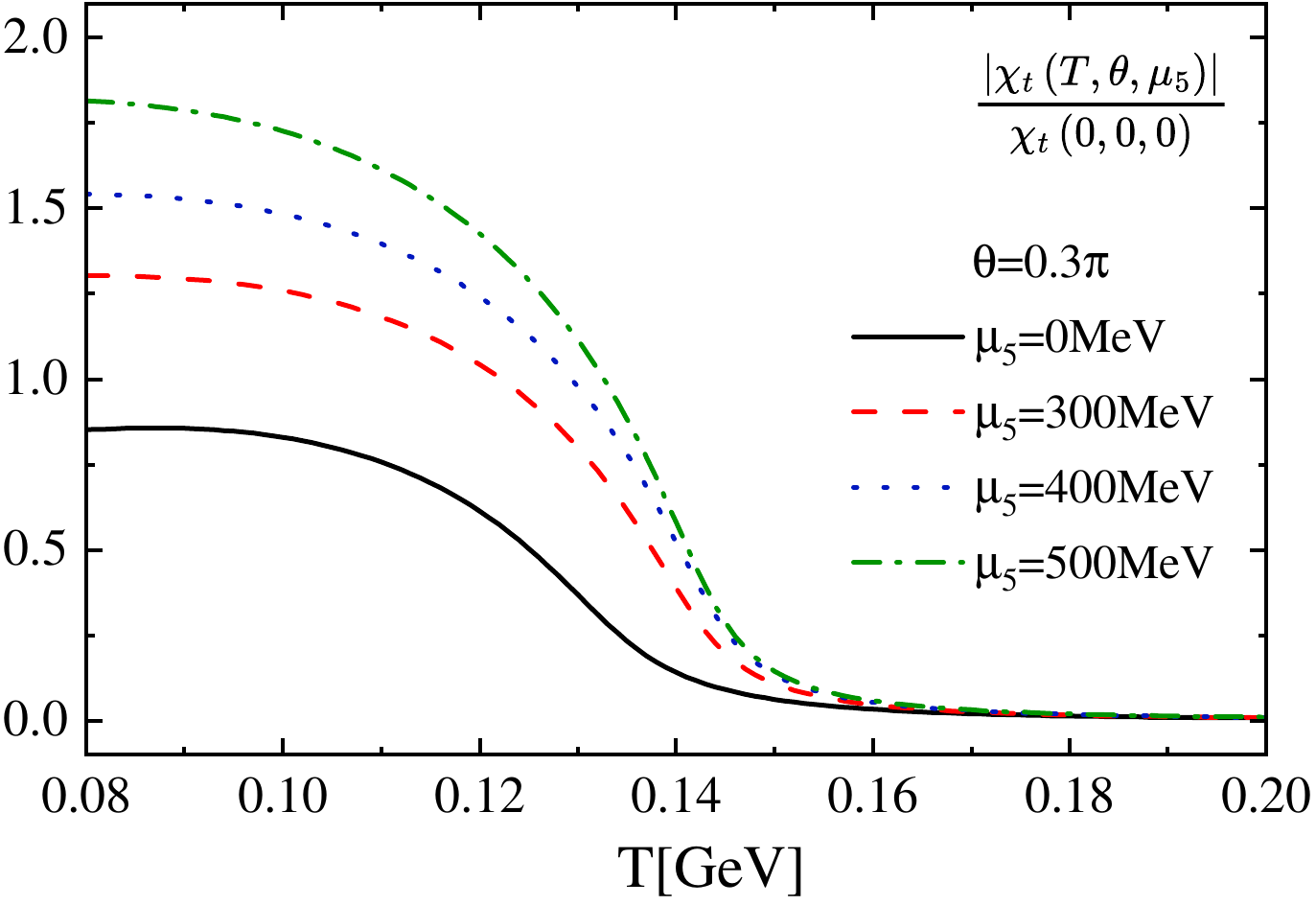}
     \subcaption{}
      \label{fg:fg2-2}
    \end{subfigure}
  \hfill  
    \begin{subfigure}{0.48\textwidth}
     \includegraphics[width=\linewidth]{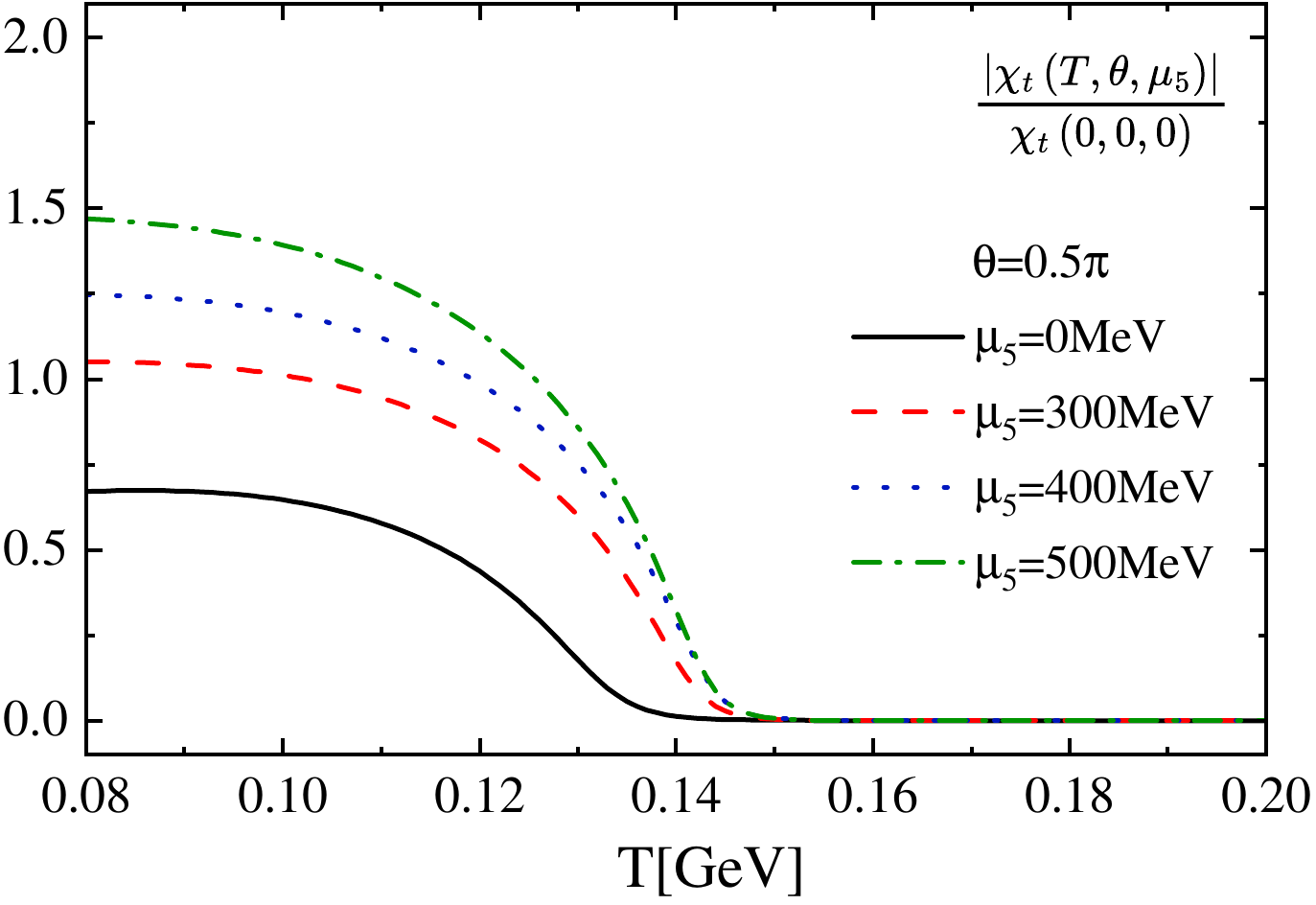}
     \subcaption{}
      \label{fg:fg2-3}
    \end{subfigure}
  \hfill
  \begin{subfigure}{0.48\textwidth}
     \includegraphics[width=\linewidth]{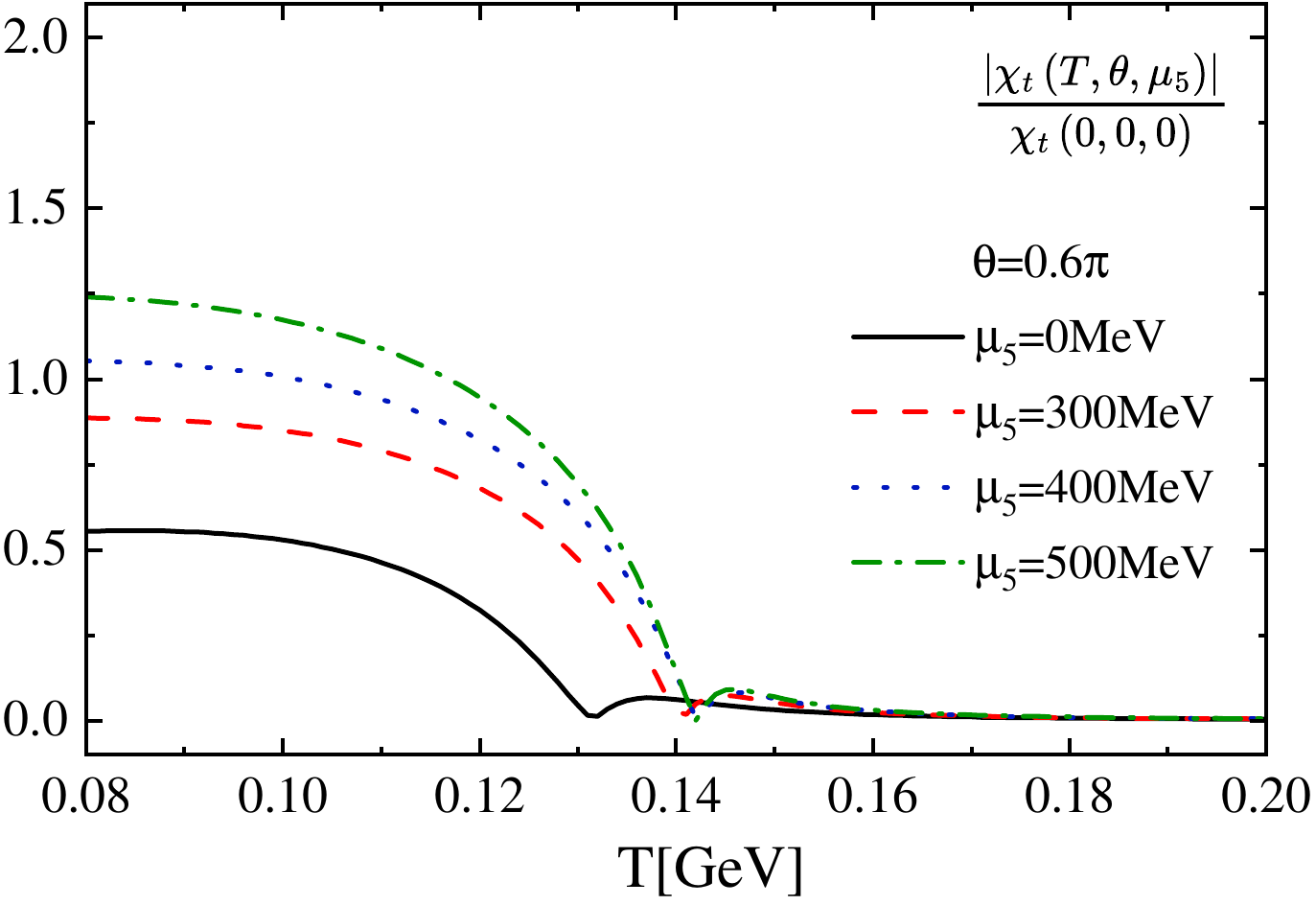}
     \subcaption{}
      \label{fg:fg2-4}
    \end{subfigure}
  \hfill  
    \begin{subfigure}{0.48\textwidth}
     \includegraphics[width=\linewidth]{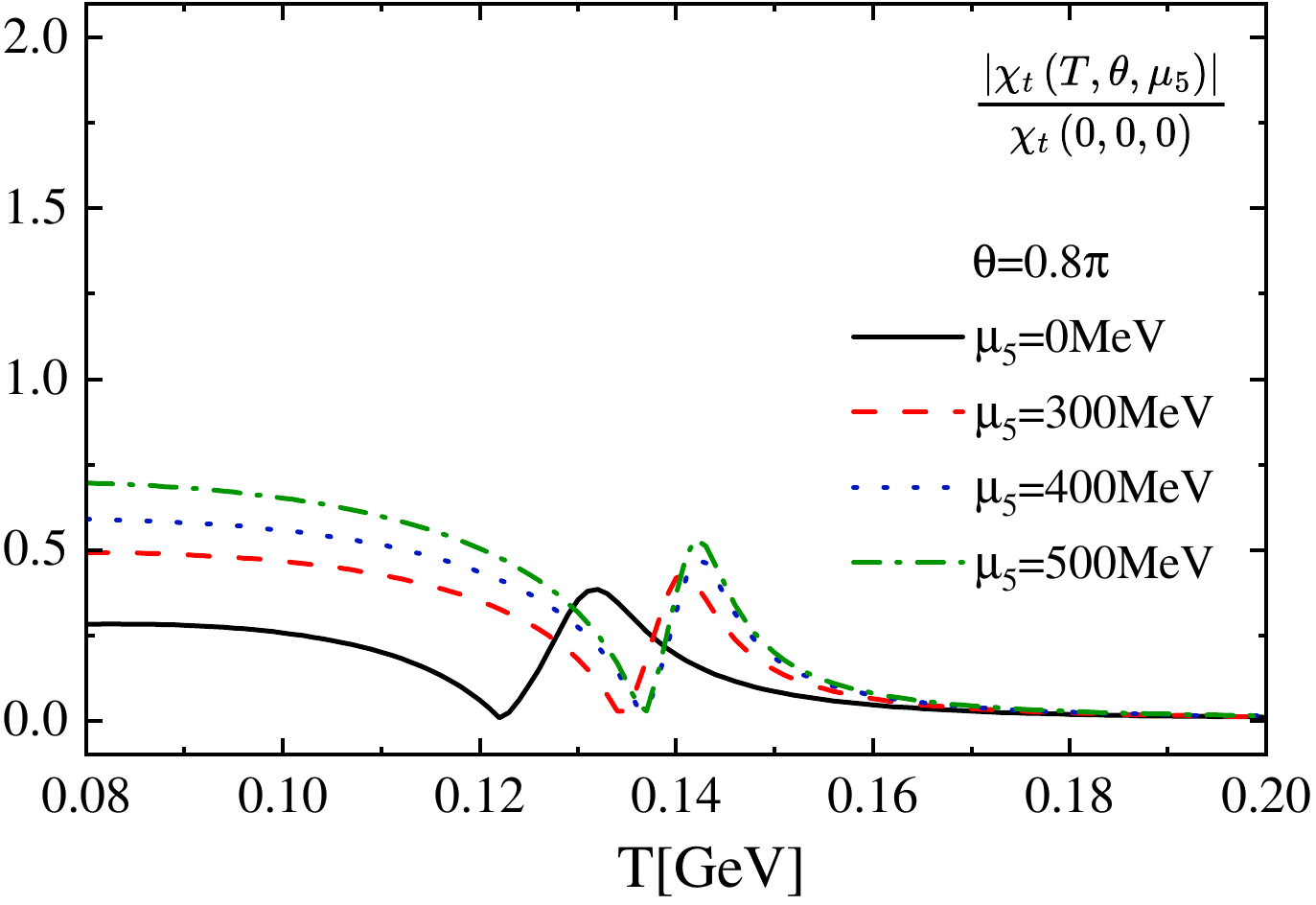}
    \subcaption{}
     \label{fg:fg2-5}
    \end{subfigure} 
  \hfill  
    \begin{subfigure}{0.48\textwidth}
     \includegraphics[width=\linewidth]{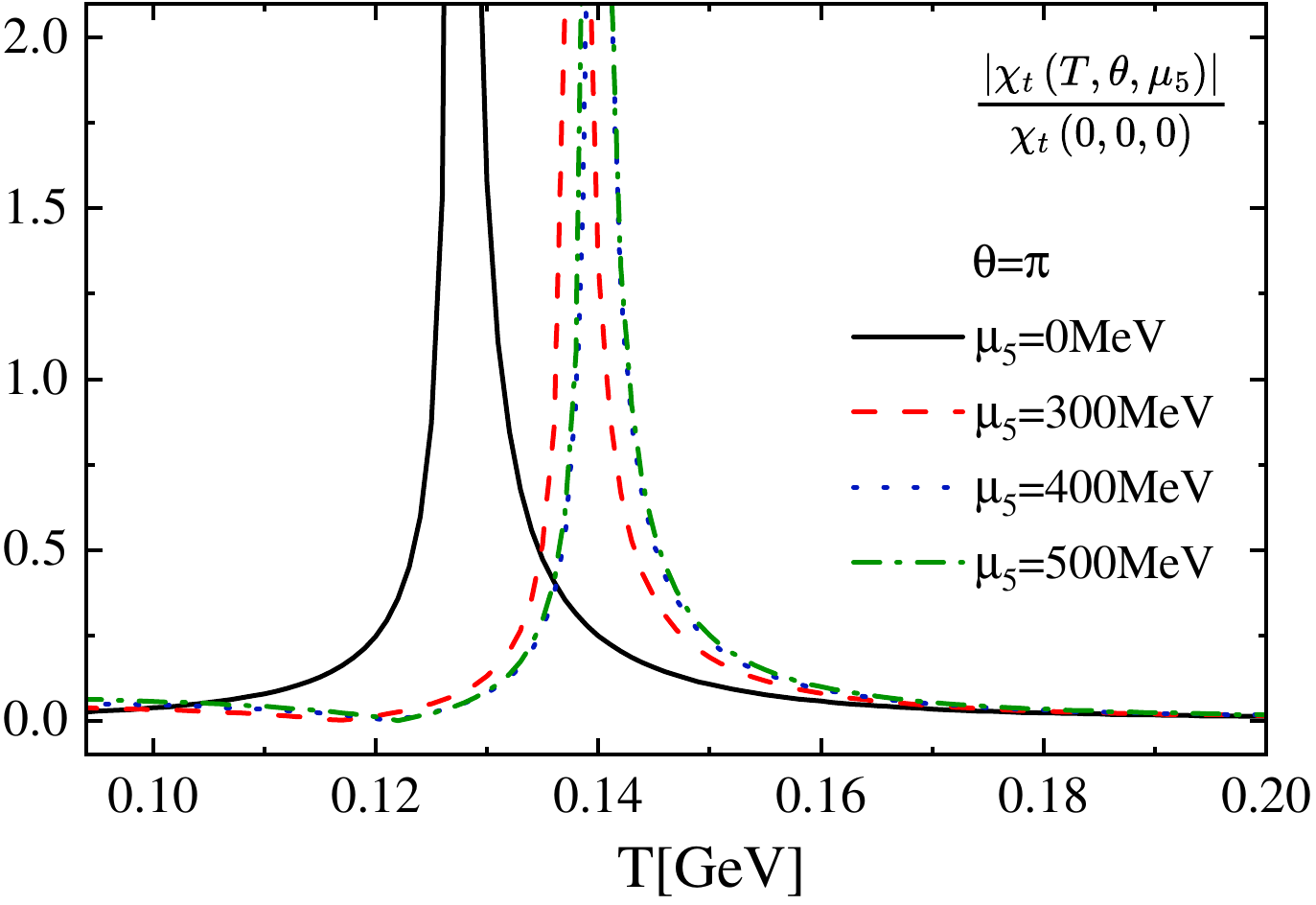}
     \subcaption{}
      \label{fg:fg2-6}
    \end{subfigure}
	
	\caption{ Absolute value of the $\theta$-dependent topological susceptibility $\chi_{\text{t}}$ versus temperature under the same conditions as that in Fig. \ref{fg:condensates1}.}
	\label{fg:topsus1}
\end{figure*}

\begin{figure*}[!htbp]
	\centering
	\begin{subfigure}{0.49\textwidth}
     \includegraphics[width=\linewidth]{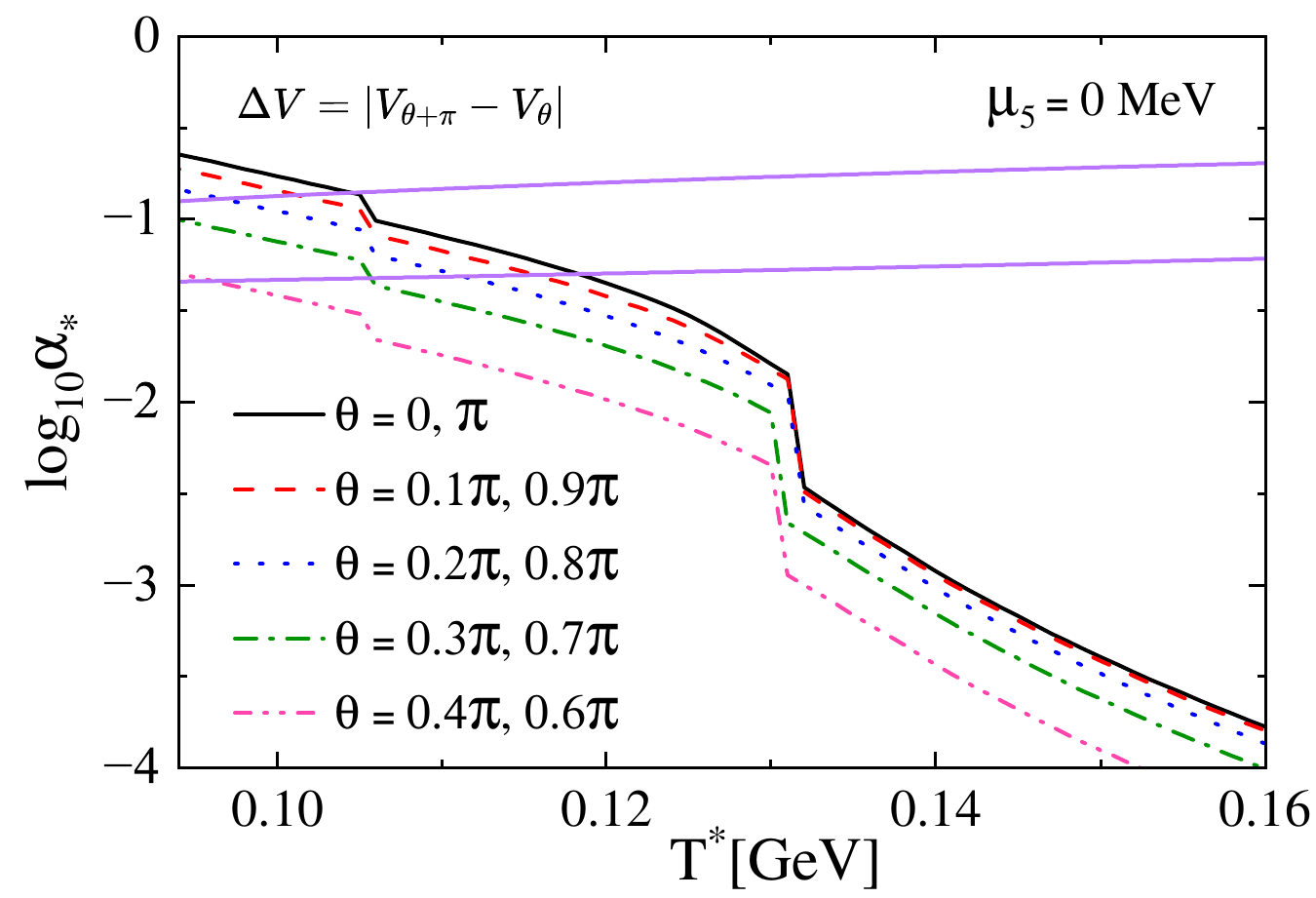}
     \subcaption{}
     \label{fg:mu0-1}
     \end{subfigure}
  \hfill
     \begin{subfigure}{0.49\textwidth}
     \includegraphics[width=\linewidth]{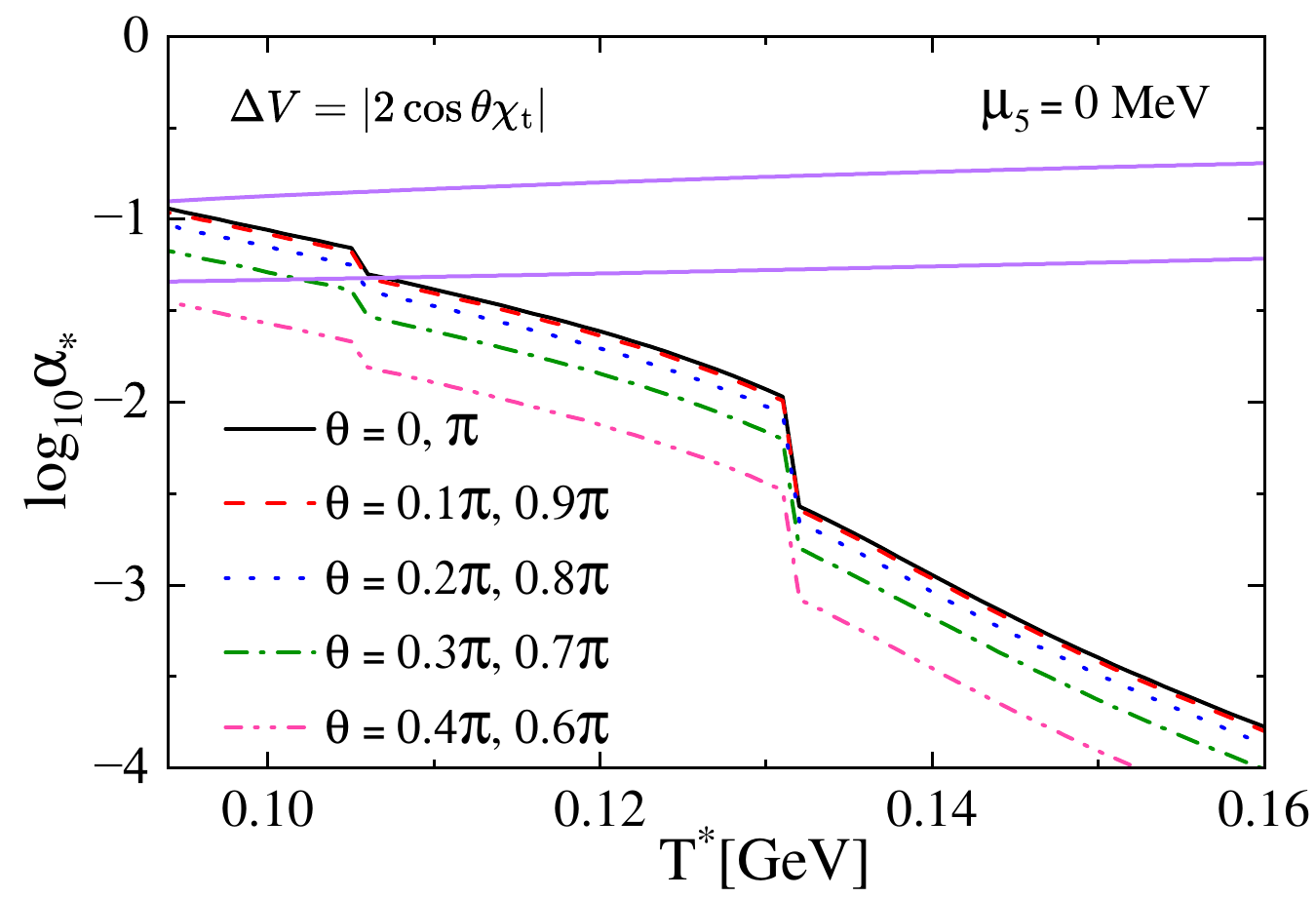}
     \subcaption{}
     \label{fg:mu0-2}
    \end{subfigure}
  	\caption{The signal strength $\alpha_*(T_*,\theta,\mu_5)$ versus the axionic DM annihilation temperature $T_*$ for different $\theta$ at $\mu_5 = 0~\text{MeV}$.
  The left panel is obtained by calculating the potential difference $\Delta V$ within the NNJL model; the right one is obtained from the cosine potential
  via evaluation of the topological susceptibility $\chi_t$. The two purple lines in both panels correspond to the 2$\sigma$ contour from Ref. \cite{NANOGrav:2023hvm}, with 
  the region inside denoting the allowed parameter space for the nHz GWs. }
	\label{fg:mu0}
\end{figure*}

\subsection{Evaluation of topological susceptibility in NNJL}

The generalized QCD topological susceptibility at finite $\theta$ \cite{Huang:2024nbd} is defined as 
\begin{eqnarray}
\chi_{\rm t}(\theta) = \frac{1}{V} \langle Q^2 \rangle &=& \int d^3x \, \langle q(0) q(x) \rangle_{\theta} \nonumber\\
                                               &=& \frac{d^2 \Omega_{\text{QCD}}}{d \theta^2}\Big|_{\theta}. 
\end{eqnarray}
Note that $\chi_t(\theta=0)$ corresponds to the conventional topological susceptibility. 

In the NNJL formalism, the generalized topological susceptibility $\chi_{\rm t}(\theta)$ can be evaluated via
\begin{eqnarray}
\chi_{\rm t}(\theta) = \frac{d^2 V}{d \theta^2}\Big|_{\theta}, 
\end{eqnarray}
where $V$ is an effective potential \footnote{It corresponds to the potential of QCD axion if $\theta$ is replaced by the axion field $a/f_a$.} 
defined as 
\begin{equation}
V(T,\theta,\mu_5)=\Omega_{\text{M}}(x_i(T,\theta,\mu_5),T,\theta,\mu_5), \label{eq:AP}
\end{equation} 
and $x_i(T,\theta,\mu_5)$ refer to the physical values of $\sigma'_0$ and $\eta'_0$, which are solutions of the gap equations \eqref{eq:GapEqs} at 
given $T$, $\theta$, and $\mu_5$.  Since the physical condensates are all implicitly dependent on $\theta$, the total differential of $V(\theta)$ 
with respect to $\theta$ satisfies the following relation
\begin{equation}
 \frac{\mathrm{d}{V}}{\mathrm{d} \theta}=\frac{\partial {V}}{\partial \theta}+\frac{\partial {V}}{\partial \sigma'_0 }\frac{\partial \sigma'_0 }{\partial \theta}
 +\frac{\partial {V}}{\partial \eta'_0}\frac{\partial \eta'_0 }{\partial \theta} . \label{eq:diffVa}
\end{equation}
Therefore to calculate $\chi_{\rm t}$, we need to compute the 1-2th partial derivatives of each of the physical condensates 
with respect to $\theta$,  namely $\frac{\partial^{(n)}x_i}{\partial \theta^{(n)}}|_{\theta}$, where $n=1,2$. 
This can be fulfilled by taking the successive derivatives of the gap equations \eqref{eq:GapEqs} with respect to $\theta$ . 
First, the system of linear equations with $\frac{\partial x_i}{\partial \theta }$ as variables
can be obtained by differentiating the gap equations \eqref{eq:GapEqs} with respect of $\theta$ and thus $\frac{\partial x_i}{\partial \theta}|_{\theta}$ 
can be calculated using the numerical solutions of \eqref{eq:GapEqs} at $\theta$. Second, the linear equations with $\frac{\partial^{(2)}x_i}{\partial \theta^{(2)}}$ 
as variables can be obtained by taking the second-order derivative of \eqref{eq:GapEqs} with respect to $\theta$ and then 
$\frac{\partial^{(2)}x_i}{\partial \theta^{(2)}}|_{\theta}$ can be computed (since $x_i|_{\theta}$ and $\frac{\partial x_i}{\partial \theta}|_{\theta}$ 
are all known). Such a method has been used to evaluate the mass and self-coupling of QCD axion within the NJL framework \cite{Abhishek:2020pjg,Zhang:2025lan}.

\begin{figure}[!htbp]
	\centering
	 \includegraphics[width=\linewidth]{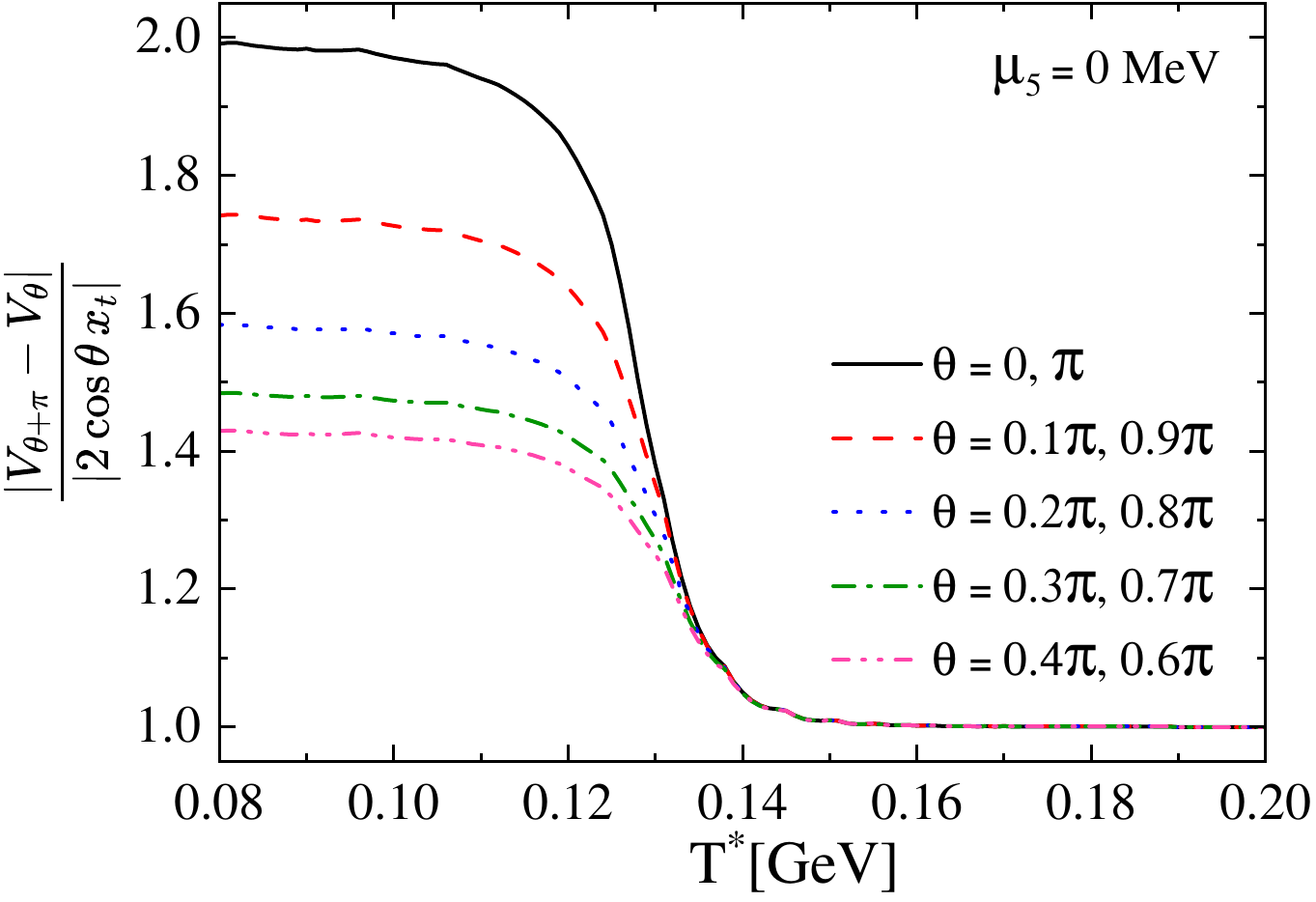}
  	\caption{The ratio of the potential bias calculated directly within the NNJL model to that from the simple cosine potential versus the axionic DM annihilation 
  temperature $T_*$ for different $\theta$ at zero $\mu_5$.}
	\label{fg:mu0ratio}
\end{figure}

\section{Numerical results and discussion}  \label{sec:results}

We show the numerical results obtained within the NNJL model in this section. We mainly concentrate on the situation with a relatively 
larger $\mu_5$ (we assume $\mu_5>0$), which may exert a significant impact on the QCD bias. 

\begin{figure}[!htbp]
\centering
\includegraphics[width=\linewidth]{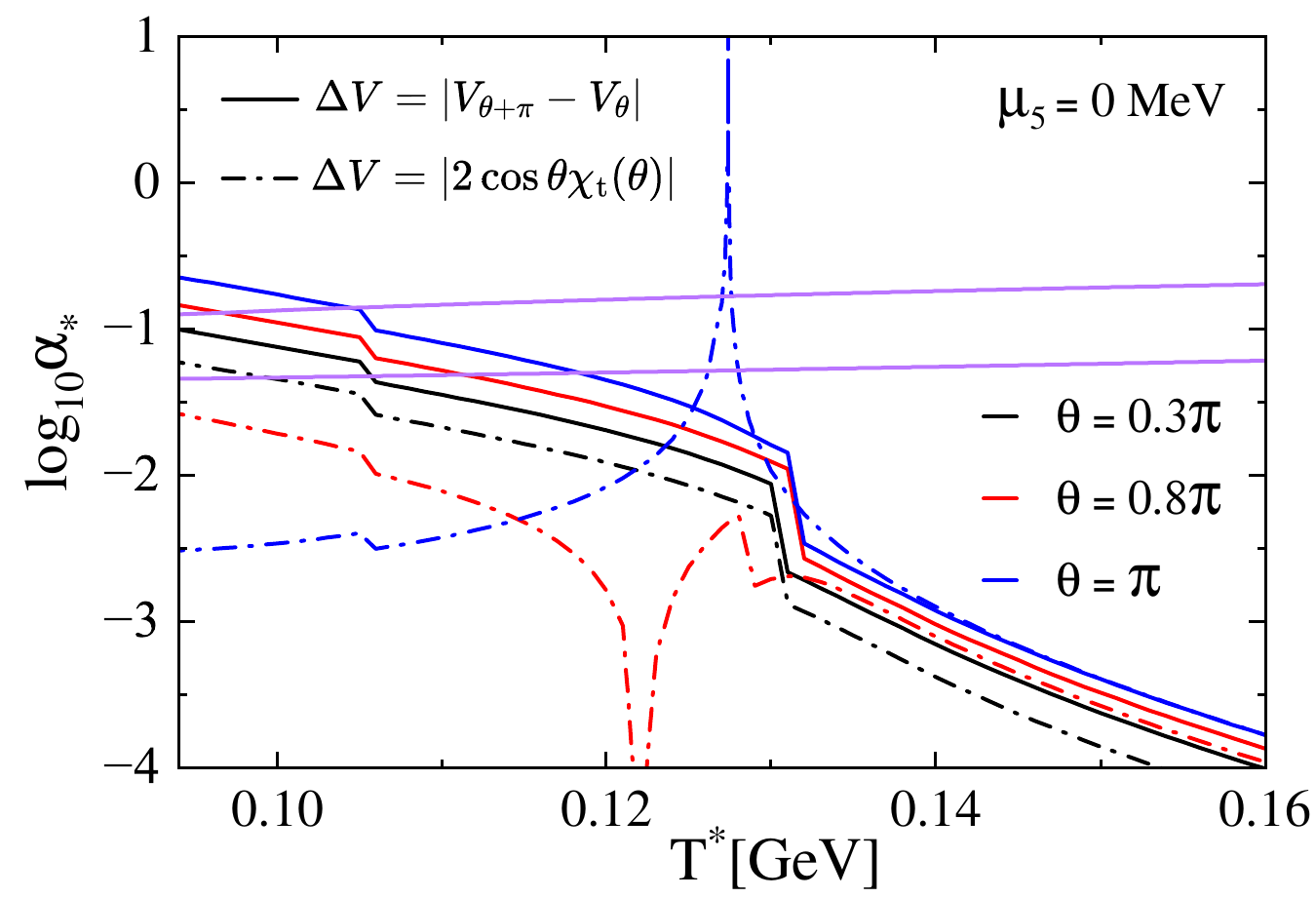}
\caption{
Comparison of the signal strength $\alpha_*(T_*,\theta,\mu_5)$ versus $T_*$ obtained from the potential bias within the NNJL model (solid lines) and those from 
the \(\theta\)-dependent topological susceptibility $\chi_t(\theta)$ (dash-dotted lines), for different $\theta$ at $\mu_5 = 0~\text{MeV}$. The two purple lines are 
same as that in Fig. \ref{fg:mu0}. }
\label{fg:NJLvsGDM0}
\end{figure}

The normalized condensates $\langle \bar{Q}Q \rangle$ and $\langle \bar{Q}i\gamma_5Q \rangle$ as functions of $T$ are shown in 
Fig. \ref{fg:condensates1} for various $\theta \in [0,\pi]$ with fixed values of $\mu_5 =0, 300, 400, 500~\text{MeV}$. Panel (a) 
displays that for zero $\theta$, the scalar condensate |$\langle \bar{Q}Q \rangle$| increases with $\mu_5$ 
\footnote{Consistent with the lattice results, there is no the so called inverse catalytic effect near the crossover temperature obtained in the local NJL model. }, 
which is consistent with Refs. \cite{Ruggieri:2020qtq,Ruggieri:2016ejz}. Panels (b)-(f) demonstrate that for nonzero $\theta$, not only 
|$\langle \bar{Q}Q \rangle$|, but also the pseudo-scalar condensate |$\langle \bar{Q}i\gamma_5Q \rangle$| grows with $\mu_5$. 
Panel (f) illustrates Danshen's phenomenon at $\theta=\pi$: the condensate $\langle \bar{Q}i\gamma_5Q \rangle$ exhibits two degenerate 
solutions with opposite signs at low temperature, signaling the spontaneous CP symmetry breaking. We see that the CP restoration temperature 
also increases with $\mu_5$ due to the catalytic effect of chirality imbalance.

The magnitude of the normalized $\chi_t(T,\theta,\mu_5)$ versus $T$ under the same conditions as in Fig. \ref{fg:condensates1} is 
demonstrated in Fig. \ref{fg:topsus1}. Panel (a) displays that the conventional topological susceptibility decreases with T for fixed 
$\mu_5$ but increases with $\mu_5$ for fixed $T$. We see that at low temperatures, the  $\chi_t({\theta = 0})$ at $\mu_5=500~\text{MeV}$ is roughly 
twice that at zero $\mu_5$. The generalized topological susceptibility $\chi_t({\theta \neq 0})$ exhibits similar behavior for 
$\theta\leq 0.5\pi$, as shown in panels (b)-(c). Panels (a)-(c) indicate that for small values of $\theta$, $\chi_t(\theta)$ monotonically 
decreases with $\theta$. Panels (d)-(e) show that for large angles $\theta=0.6\pi$ and $0.8\pi$, the absolute value of $\chi_t(\theta)$ first 
falls to zero with increasing $T$, then increases to a local maximum and gradually declines thereafter. This abnormal behavior can be 
understood using the $\chi_t(\theta)$ decomposition \cite{Huang:2024nbd} 
\footnote{There is a typo in Eq. (4) of Ref. \cite{Huang:2024nbd}, where the imaginary unit $i$ should be replaced with a minus sign.}. 
\begin{equation}
\chi_t(T,\theta,\mu_5)=-\frac{1}{4}m_0\langle{\bar{\psi}\psi}\rangle+\frac{1}{4}m_0^2\chi_\eta(T,\theta,\mu_5),
\label{eq:sus1+1}
\end{equation}
where
\begin{equation}
\chi_\eta(T,\theta,\mu_5)=\int_T dx^4 \langle(\bar{\psi}i\gamma_5\psi)(\bar{\psi}(x)i\gamma_5\psi(x))\rangle_{\theta,\mu_5}.
\end{equation}
For large enough $\theta$, the second term in \eqref{eq:sus1+1} becomes dominant over the first one and a $\chi_t$ peak develops 
due to the rapid decrease of the pseudoscalar condensate with $T$ since $\chi_\eta \sim \frac{\partial}{\partial T} \langle\bar{\psi}i\gamma_5\psi\rangle $. 
This phenomenon becomes more pronounced for $\theta=\pi$ where a second order phase transition for CP restoration happens 
at $T=T_c$ where |$\chi_t$| diverges, as demonstrated in Panel (f). This implies that |$\chi_t(\theta)$| can become quite
large for $T \sim T_c$ at $\theta=\pi$ or $\theta \sim \pi $ at $T=T_c$. The properties of $\chi_t(\theta)$ are crucial for understanding 
the GW signal strength obtained via Method II and Method III.  

\begin{figure*}[!htbp]
	\centering
	\begin{subfigure}{0.49\linewidth}
		\includegraphics[width=\linewidth]{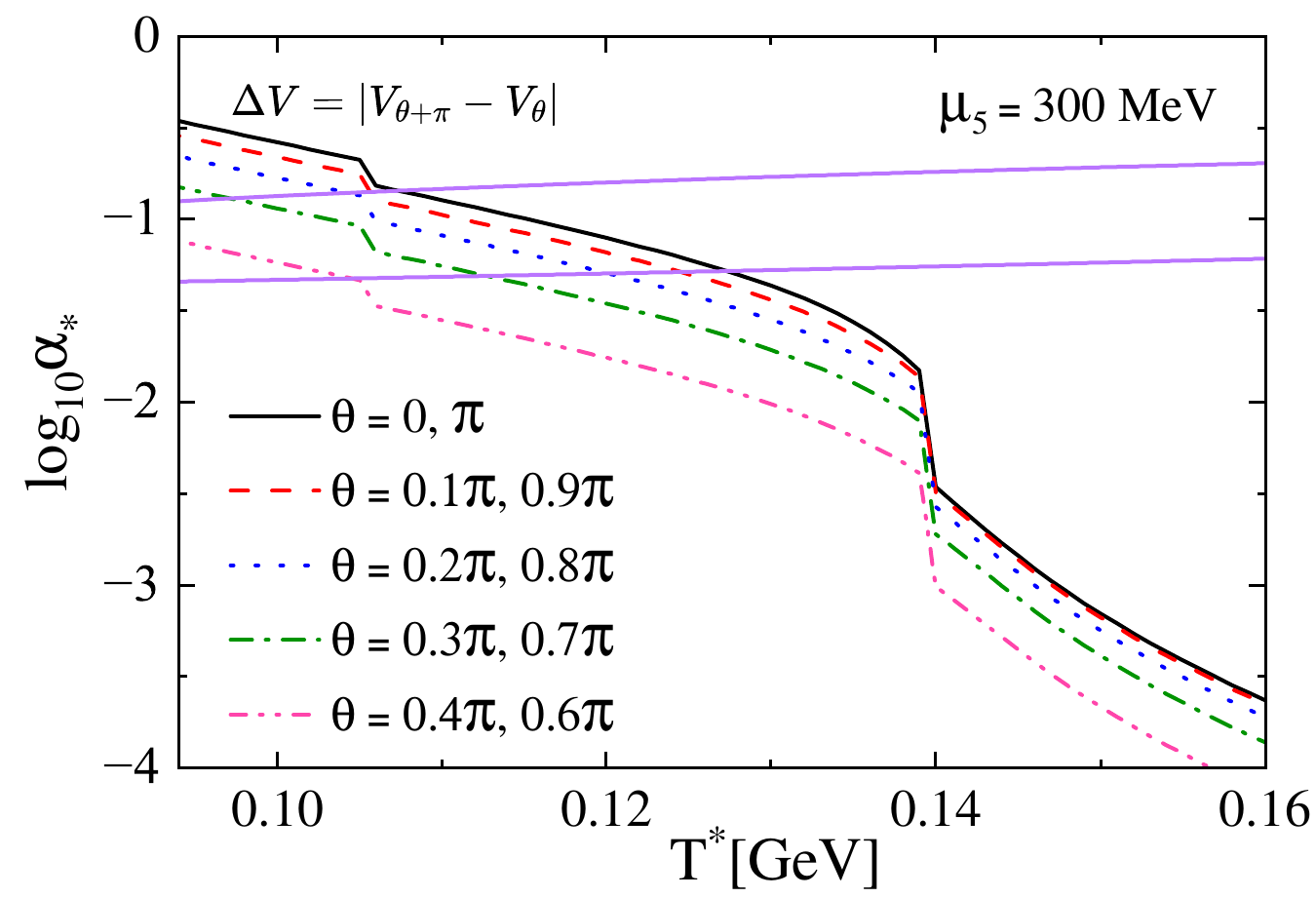}
	\subcaption{}	
	\end{subfigure}
	\hfill
	\begin{subfigure}{0.49\linewidth}
		\includegraphics[width=\linewidth]{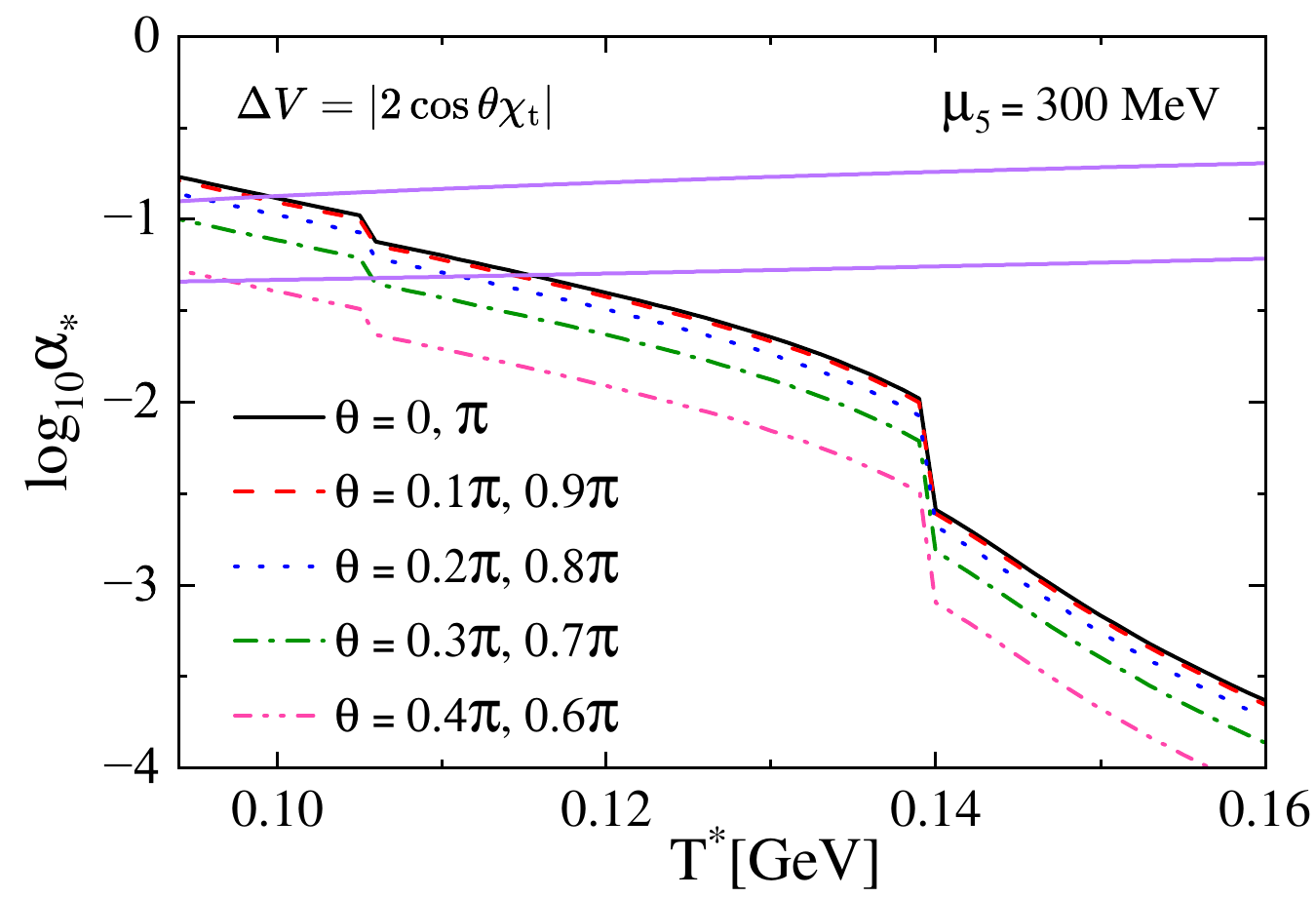}
	\subcaption{}	
	\end{subfigure}
	\caption{Same as in Fig. \ref{fg:mu0}, but for $\mu_5 = 300~\text{MeV}$.}
	\label{fg:mu300}
\end{figure*}

\begin{figure*}[!htbp]
	\centering
	
	\begin{minipage}{0.49\linewidth}
		\centering
		\includegraphics[width=\linewidth]{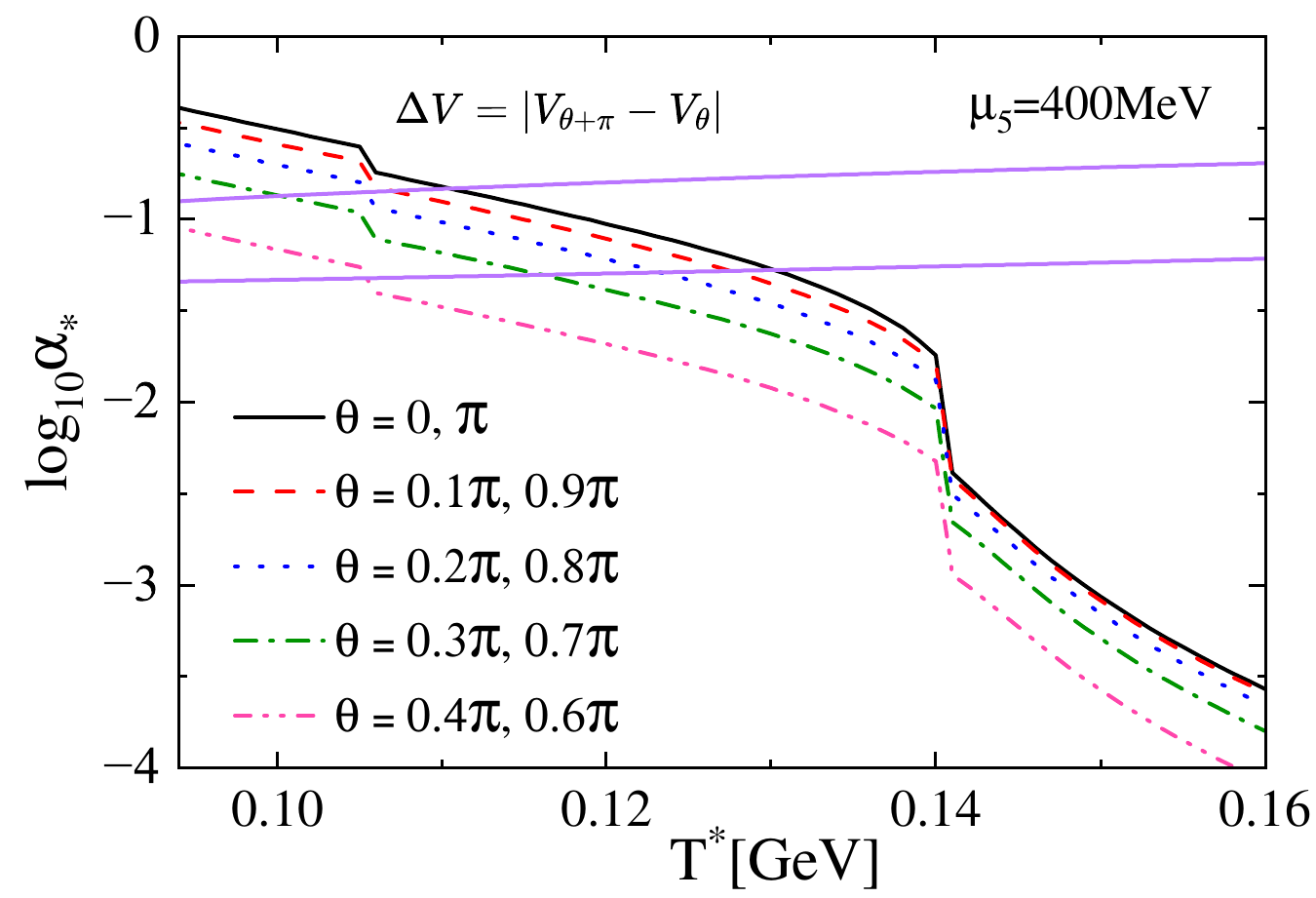}
		
		(a)
	\end{minipage}
	\hfill
	\begin{minipage}{0.49\linewidth}
		\centering
		\includegraphics[width=\linewidth]{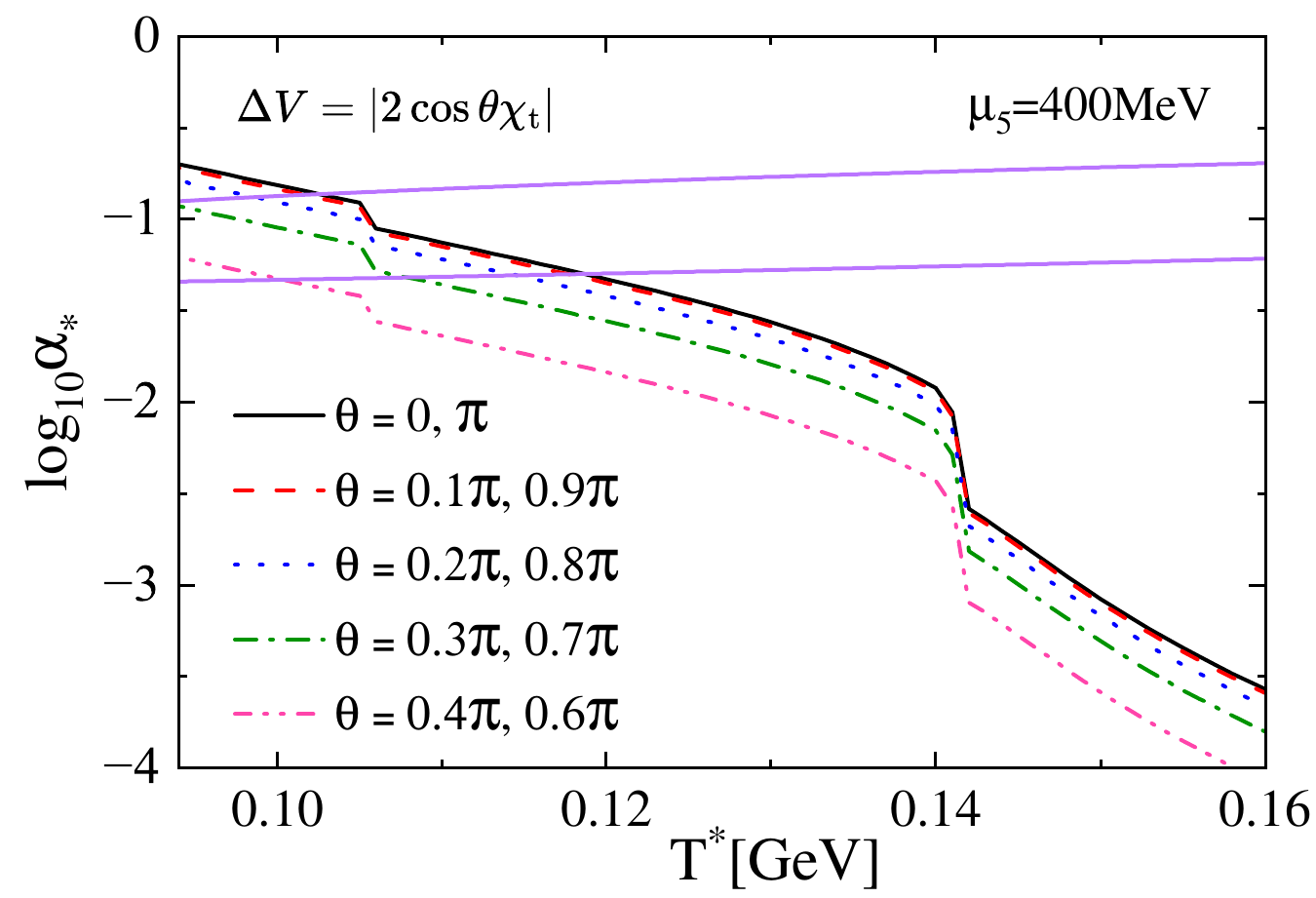}
		
		(b)
	\end{minipage}
	
	\caption{Same as in Fig. \ref{fg:mu0}, but for $\mu_5 = 400~\text{MeV}$.
} 
	\label{fg:mu400}
\end{figure*}	

\begin{figure*}[!htbp]
	\centering
	\begin{minipage}{0.49\linewidth}
		\centering
		\includegraphics[width=\linewidth]{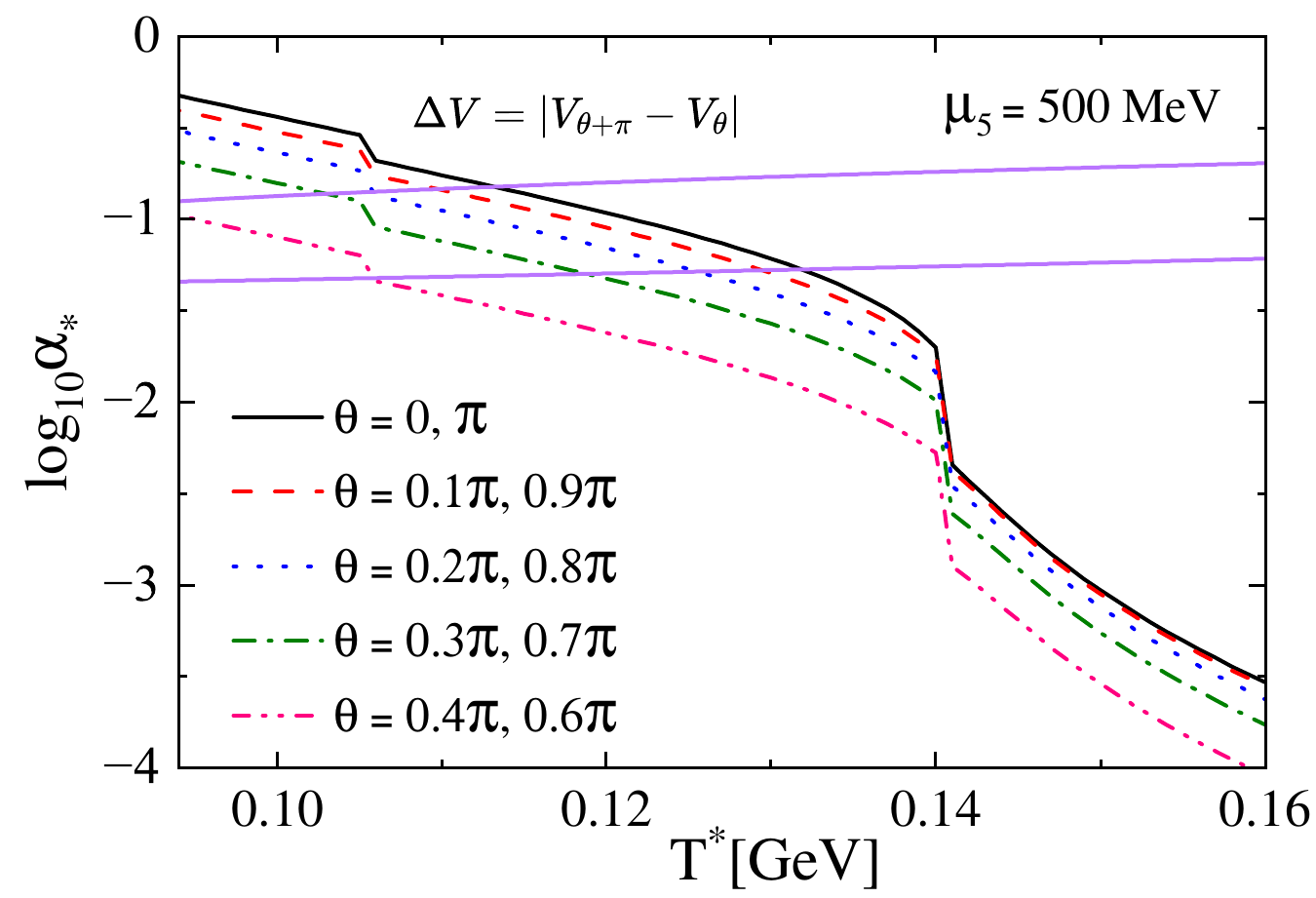}
		
		(a)
	\end{minipage}
	\hfill
	\begin{minipage}{0.49\linewidth}
		\centering
		\includegraphics[width=\linewidth]{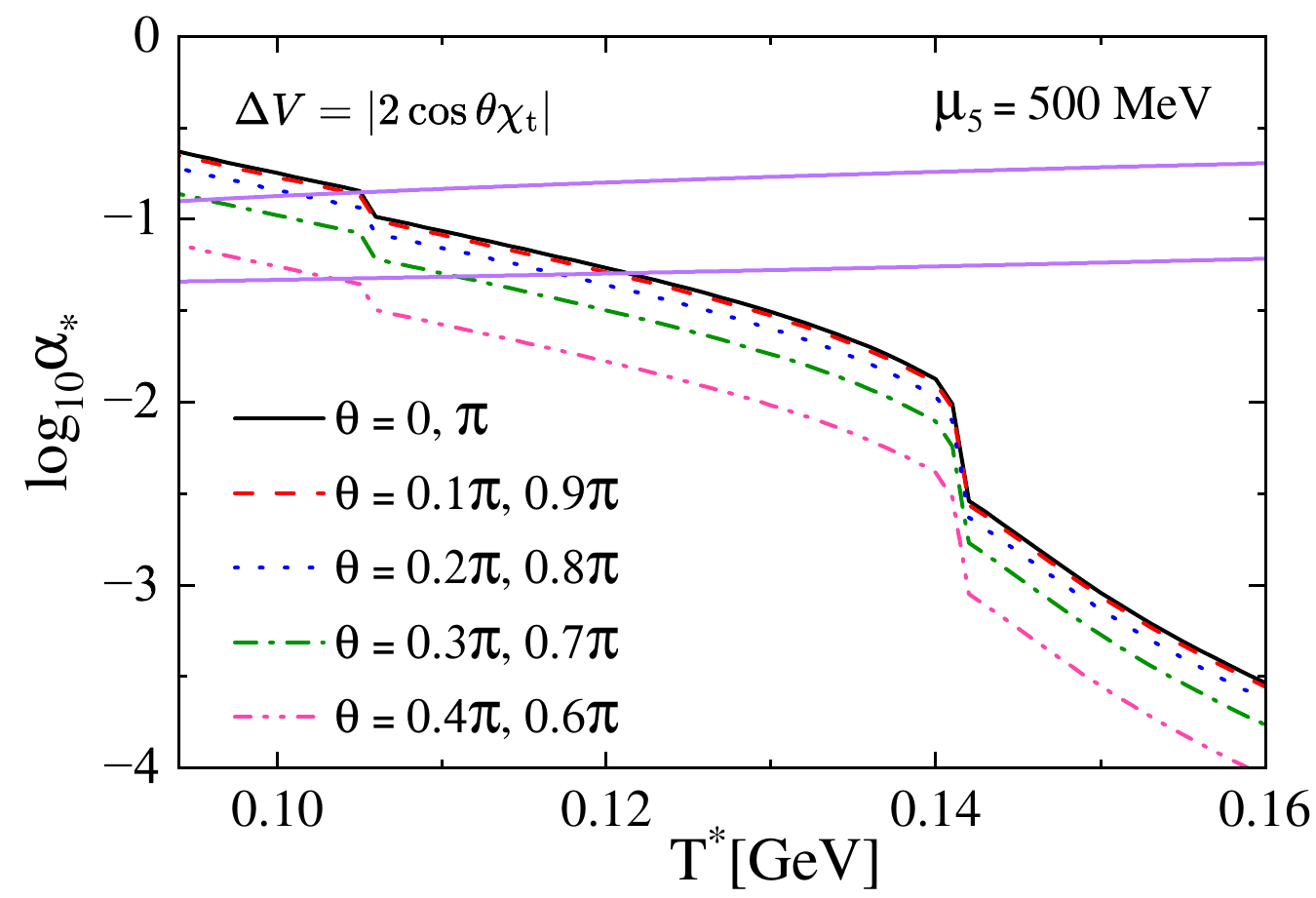}
		
		(b)
	\end{minipage}
	
	\caption{Same as in Fig. \ref{fg:mu0}, but for $\mu_5 = 500~\text{MeV}$.
}
	\label{fg:mu500}
\end{figure*}

Figures \ref{fg:mu0} shows the GW signal strength versus the axionic DM annihilation temperature $T_*$ for different $\theta$ angles 
at vanishing $\mu_5$. The left and right panels are obtained via Method I and Method II, respectively. Two purple lines in these panels 
correspond to the 2$\sigma$ contour from Ref.\cite{NANOGrav:2023hvm}, with the region inside denoting the allowed parameter space. 
The wiggles in the signal strength curves arise from the thresholds encoded in $g_*(T)$. Fig. \ref{fg:mu0} demonstrates $\alpha_*(\theta)=\alpha_*(\pi-\theta)$, 
which justifies relation \eqref{eq:same2angles}. This means that the GW signal strength at $\theta=0$ is identical to that at $\theta=\pi$, 
which is quite different from that obtained in Ref. \cite{Huang:2024nbd}. Both panels show that, for a fixed $\theta$, the signal strength 
drops with $T$. In Method II, this can be explained qualitatively by the reduction of $\chi_t(\theta=0)$ with $T$, as shown in Fig. \ref{fg:fg2-1} and implied  
by the cosine potential. On the other hand, for a fixed temperature, the signal strength decreases with $\theta$ in the interval $0\leq\theta\leq \pi/2$ 
(thus increases with $\theta$ in the interval $\pi/2\leq\theta\leq \pi$) \footnote{The logarithm of the signal strength at $\theta=\pi/2$ 
is minus infinity since $\Delta V (\theta=\pi/2) = 0$.}. The left panel indicates that the small angle range $0\leq\theta\leq 0.4\pi$ 
and the corresponding large supplementary angle range $0.6\pi\leq\theta\leq \pi$ both fall into the two-line regime. The allowed T-range 
for the nHz GW is near $100 \,\text{MeV}$ but below the chiral pseudo-critical temperature $T_{pc}$, which shifts toward lower (higher) 
temperatures with increasing $\theta$ for $0\leq\theta\leq 0.4\pi$ ($0.6\pi\leq\theta\leq \pi$). Qualitatively similar results are 
observed in the right panel, where the T-range entering the two-line region shifts toward lower temperatures. 

\begin{figure}[!htbp]
	\centering
	\begin{minipage}{0.99\linewidth}
		\centering
		\includegraphics[width=\linewidth]{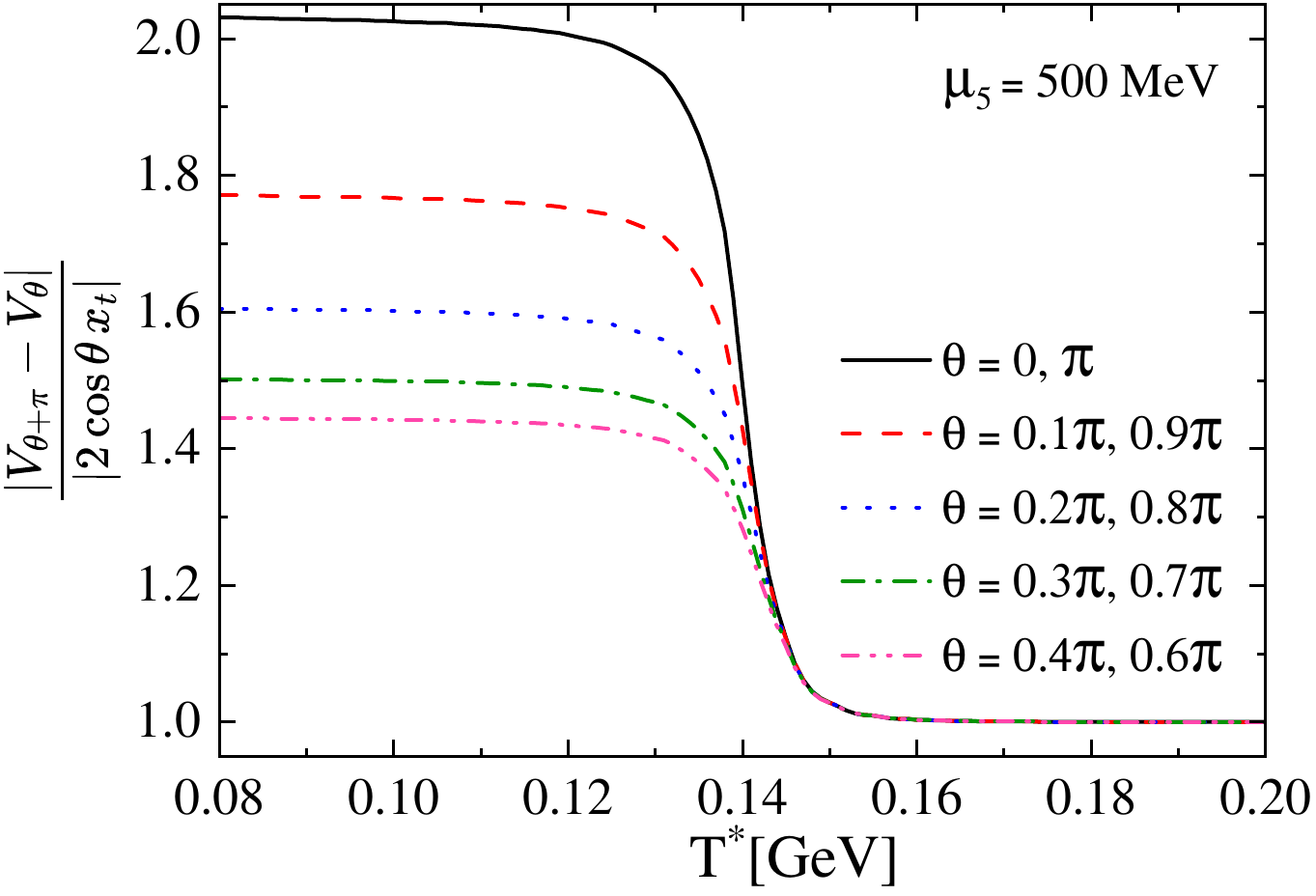}
	\end{minipage}
\caption{Same as in Fig. \ref{fg:mu0ratio}, but for $\mu_5 = 500~\text{MeV}$.
}
	\label{fg:mu500ratio}
\end{figure}	

Figure. \ref{fg:mu0ratio} demonstrates the ratio of the potential bias calculated using Method I to that computed via Method II 
for various values of $\theta$ at vanishing $\mu_5$. We see that for $T \lesssim T_{pc}$, the potential bias obtained from Method II 
is significantly smaller than that from Method I. This explains why, in contrast to Fig. \ref{fg:mu0-1}, the temperature window for the nHz GW 
shifts toward lower temperatures in Fig. \ref{fg:mu0-2}. Note that both methods yield nearly the identical potential bias for higher 
temperatures, implying that the cosine potential provides a good approximation for $T>1.1 \sim 1.2T_{pc}$.

The comparison between the GW signal strength computed using Method I and those obtained via Method III for several values of $\theta$ 
at vanishing $\mu_5$ is displayed in Fig. \ref{fg:NJLvsGDM0}. The discrepancies in conclusions between the two methods are pronounced. 
First, the general relation \eqref{eq:same2angles}, which holds for Method I (also for Method II), breaks down for Method III. 
For instance, when using Method III, the resulting GW signal strength at $\theta=0$ differs substantially from that at $\theta=\pi$.  
Second, for $\theta$ substantially smaller than $\pi$, the GW signal strength evaluated via $\chi_t(\theta)$ is noticeably   
smaller than that computed using Method I or Method II under the same conditions. This can be attributed to the suppression of $\chi_t(\theta)$ 
with increasing $\theta$, as shown in Fig. \ref{fg:topsus1}. For $\theta=\pi$,  we confirm the criticality found in \cite{Huang:2024nbd}: 
due to the thermal phase transition for the CP restoration, the signal strength obtained from $\chi_t(\theta)$ exhibits a critical peak 
at $T \sim T_c$. Unlike the local NJL result in Ref. \cite{Huang:2024nbd}, the peak obtained using Method III within the NNJL model exhibits 
a finite width. This means a narrow $\theta$ range near $\pi$ around $T \sim T_c$ can also fall within the two-line region. Such a 
possibility should be ruled out, because the enhanced signal strength for $\theta \sim \pi$ and $T \sim T_c$ is an artifact arising 
from the incorrect use of $\chi_t(\theta)$ to evaluate the QCD potential bias.    

\begin{figure}[!htbp]
\centering
\includegraphics[width=\linewidth]{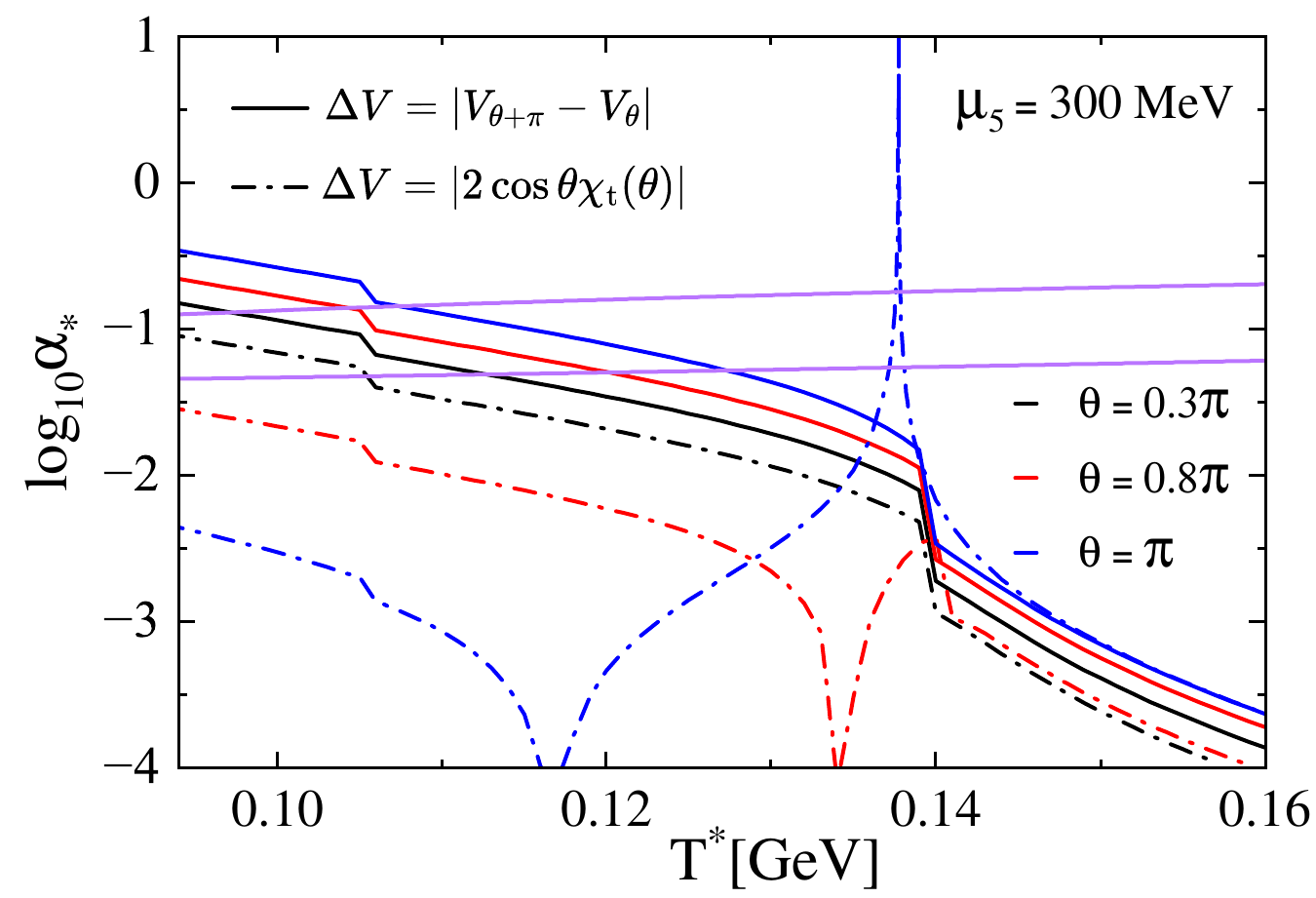}
\caption{Same as in Fig. \ref{fg:NJLvsGDM0}, but for $\mu_5 = 300~\text{MeV}$. 
}
\label{fg:NJLvsGDM300}
\end{figure}

\begin{figure}[!htbp]
	\centering
	\begin{minipage}{0.99\linewidth}
		\centering
		\includegraphics[width=\linewidth]{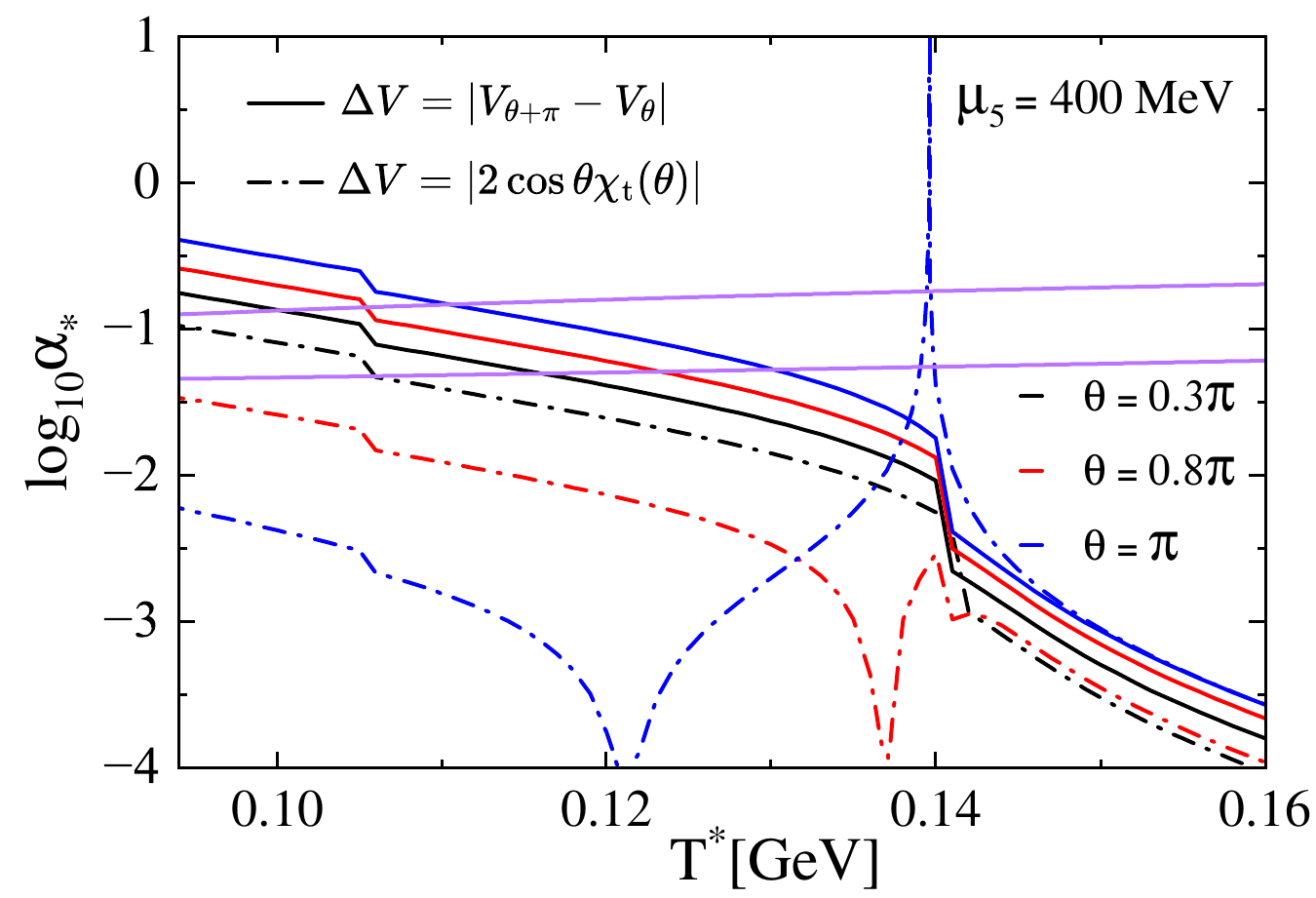}
	\end{minipage}
	\caption{Same as in Fig. \ref{fg:NJLvsGDM0}, but for $\mu_5 = 400~\text{MeV}$. }
	\label{fg:NJLvsGDM400}
\end{figure}	

\begin{figure}[!htbp]
	\centering
	\begin{minipage}{0.99\linewidth}
		\centering
		\includegraphics[width=\linewidth]{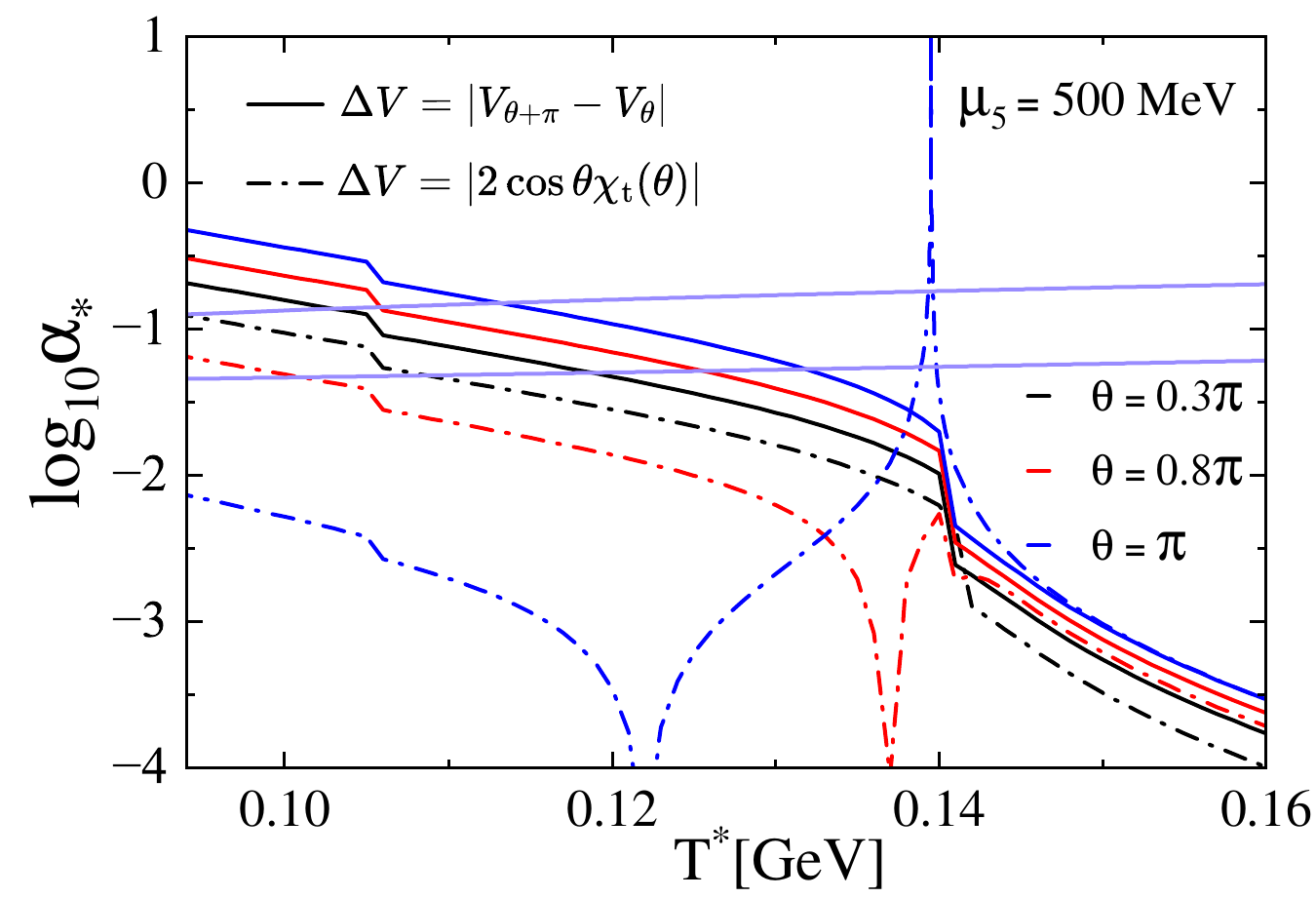}
	\end{minipage}
	\caption{Same as in Fig. \ref{fg:NJLvsGDM0}, but for $\mu_5 = 500~\text{MeV}$.}
	\label{fg:NJLvsGDM500}
\end{figure}

The GW signal strength versus the temperature calculated using Method I (II) for $\mu_5 = 300$, $400$, $500\,\text{MeV}$ and various  
$\theta$ is shown in left (right) panels of Figs. \ref{fg:mu300}-\ref{fg:mu500}, respectively. Same as in Fig. \ref{fg:mu0}, the GW 
signal strength evaluated at $\theta$ equals that at the angle $\pi-\theta$. For fixed $\theta$ and $\mu_5$, the GW signal strength
decreases with T; for fixed $T$ and $\mu_5$, it decreases with $\theta$ over the interval $ 0 \leq \theta \leq \pi/2$. Both features 
are consistent with the behavior at vanishing $\mu_5$. For fixed $T$ and $\theta$, the signal strength increases with $\mu_5$ in 
Method II is readily understood: the topological susceptibility rises with $\mu_5$, as illustrated in Fig. \ref{fg:fg2-1}. Consequently, 
the $(T,\theta)$ region that can access the two-line region moves toward higher temperatures with $\mu_5$. We find that the allowed 
temperatures for the nHz GWs obtained via method I lie below the chiral crossover temperature $T_{pc}(\mu_5)$, which rises with 
$\mu_5$ within the NNJL model. Figs. \ref{fg:mu300}-\ref{fg:mu500} show that for nonzero $\mu_5$ under identical conditions, the 
GW signal strength obtained from Method I also substantially exceeds that from Method II at low temperatures and converges toward it 
with increasing temperature. 

Figure \ref{fg:mu500ratio} displays the T-dependence of the ratio of the QCD potential bias from Method I to that from Method II 
for $\mu_5 = 500 \,\text{MeV}$ and various $\theta$ angles. For $T<T_{pc}(500)=T_{pc}(\mu_5=500\,\text{MeV}) \simeq 140 \, \text{MeV}$, 
the ratio decreases from $\sim2$ to $\sim1.4$ as $\theta$ ranges from $0$ to $0.4\pi$ (or from $\pi$ to $0.6\pi$), while it approaches 
unity for $T> 1.1$-$1.2\,T_{pc}(500)$. Thus, at large $\mu_5$, the cosine potential also serves as a good approximation for high 
temperatures.

Figures \ref{fg:NJLvsGDM300}-\ref{fg:NJLvsGDM500} present comparisons of the obtained GW signal strengthes from Methods I 
and Method III for several values of $\theta$ and $\mu_5=300, 400, 500\,\text{MeV}$. As in Fig. \ref{fg:NJLvsGDM0}, the critical 
peaks from Method III still emerge at $\theta=\pi$ and $T \sim T_c(\mu_5 \neq 0)$  
\footnote{Lower peaks also appear at $\theta$ near $\pi$, as indicated in Fig. \ref{fg:fg2-5}.}, which differ significantly from that computed via Method I. 
Compared to Fig. \ref{fg:NJLvsGDM0}, the peak shifts toward higher temperature and become narrower with increasing $\mu_5$. 
Therefore, for nonzero $\mu_5$, a small $(T,\theta)$ region near ($T_c(\mu_5),\pi$) can also lie within the two-line region when adopting 
Method III in the NNJL formalism. As mentioned above, this is an artifact arising from the improper use of $\chi_t({\theta \neq 0})$ 
and should accordingly be discarded. 

Thus, in contrast to the previous study within the local NJL formalism \cite{Huang:2024nbd}, we find that interpreting the nHz GW generation via 
the axionic DW collapse mechanism driven by the QCD bias remains possible, provided that the local CP-odd effect induced by the QCD sphaleron 
transition is taken into account. Our study suggests that the $\theta$-dependence of the QCD potential bias is quite different from that reported 
in Ref. \cite{Huang:2024nbd}. At $\mu_5=0$,  the QCD potential bias can become sufficiently large to trigger axionic domain-wall annihilation 
and generate nHz GWs for $ T \sim 100\,\text{MeV}$, apart from a narrow angle region near $\theta=\pi/2$. When the chirality imbalance is included, 
the potential bias grows with increasing $\mu_5$, and the allowed parameter window for the nHz GWs shifts toward higher $T$ and the angle $\pi/2$. 

It is still difficult to estimate the possible magnitude of $\mu_5$ in the local hot CP-odd domain during the epoch of QCD phase transition in  
the early Universe, just as it is in the heavy-ion collisions. The recent lattice QCD study \cite{Bonanno:2023thi} suggests that the sphaleron 
rate $\Gamma_s$ of $N_f=2+1$ QCD at $T=230~\text{MeV}$ is as large as $\Gamma_s/T^4=0.310$ and the extrapolated ratio at the pseudo-critical 
temperature $T_{pc}=155~\text{MeV}$ can reach $0.7-1.0$. This marks a significant step from using theoretical estimates to having genuine, 
non-perturbative predictions for $\mu_5$ or the chiral charge density. From the energy scale viewpoint, if $\mu_5$ lies within the range 
$0<\mu_5\leq \pi T_{pc} \approx 480 \text{MeV}$ for T near $T_{pc}$, our study indicates that the $\theta$ windows for the nHz GW exist for 
both the small-$\theta$ region near zero and large-$\theta$ one near $\theta=\pi$. In such cases, a large $\mu_5$ can markedly shift 
the ($\theta,T$) parameter space for nHz GWs by counterbalancing the suppression of the QCD potential bias induced by $\theta$ 
and $T$. 

The effect of QCD deconfinement phase transition can be partially incorporated in the NNJL formalism by introducing the Polyakov-loop dynamics 
(namely the so-called PNNJL model \cite{Hell:2008cc}). For simplicity, we don't take into account the Polyakov-loop degrees of freedom in this 
work. Note that it has been demonstrated that the main conclusion obtained in the local NJL is qualitatively consistent with that from the 
Polyakov-loop extended NJL model (PNJL) in Ref. \cite{Huang:2024nbd}. Since the relation $\Delta V(\theta)=\Delta V(\pi-\theta)$ always  
holds, we may expect that including the Polyakov-loop dynamics does not qualitatively alter our main results obtained in the NNJL model.

\section{Conclusions}\label{sec:conclusion}

In this work, we study the influence of the chirality imbalance in hot local CP-odd domain due to QCD sphaleron transitions  
on the possible interpretation of nHz GWs from the QCD-biased axionic DW annihilation. The QCD potential bias near the chiral 
transition temperature is calculated directly within the nonlocal NJL model with instanton-induced interactions by introducing 
the chiral chemical potential $\mu_5$ and the $\theta$ angle. The resulting GW signal strength is then evaluated and compared 
with the NG15 data.  

For fixed $T$ and $\mu_5$, we confirm that the GW signal strength decreases (increases) with increasing $\theta$ for $0<\theta \leq \pi/2$ 
($\pi/2<\theta \leq \pi$) and both angels $\theta$ and $\pi-\theta$ yield the same potential bias. On the other hand, finite $\mu_5$ 
enhances the potential bias due to the catalysis effect. Unlike that reported in Ref. \cite{Huang:2024nbd}, we find that at zero $\mu_5$, 
the interpretation of nHz GWs generated by the QCD-biased axionic DW collapse is still possible for not only a small $\theta$ range 
near zero, but also the large $\theta$ range around $\pi$.  When the chirality imbalance is taken into account, the ($\theta,T$) parameter 
space for nHz GWs shifts toward the angle $\pi/2$ and higher temperature.

We find that for temperatures above the chiral crossover temperature, the potential bias obtained from the NNJL potential 
approaches that calculated using the simple cosine potential. In contrast, at low temperatures, the potential bias is significantly 
underestimated when the cosine potential is adopted. We point out that the modified cosine potential with the $\theta$-dependent 
topological susceptibility introduced in Ref. \cite{Huang:2024nbd} is impropriate for evaluating the QCD bias. First, this method violates 
the general relation $\Delta V (\theta)=\Delta V (\pi-\theta)$. Second, this method produces an artifact at $\theta=\pi$ 
that the GW signal strength diverges at the critical temperature for the CP symmetry restoration phase transition.  

Thus we conclude that impacts of the local CP breaking and chirality imbalance in hot QCD on the axionic DM interpretation 
of NG15 data are complicated. This work is limited to the mean field approximation (MFA) and two-flavor case. It is worthwhile to 
investigate the influence beyond the MFA and the role of the strange quark on the QCD-biased signal strength. 
Moreover, it is reported in Refs. \cite{Zhang:2012rv,Yu:2014sla} that chiral imbalance can be dynamically induced by a repulsive 
axial-vector interaction based on the instanton–anti-instanton molecule picture. It is also interesting to investigate how such 
a repulsive interaction influences the QCD-biased signal strength within the NNJL formalism.

\vspace{15pt}
\noindent{\textbf{\large{Acknowledgements}}}\\\\
This work was supported by the National Natural Science Foundation of China (NSFC) under Grant No. 11875127.

\end{document}